\documentclass[
superscriptaddress,
nofootinbib,
amsmath,amssymb,
aps,
twocolumn,
floatfix,
]{revtex4-2}

\usepackage{xcolor}
\definecolor{myGreen}{RGB}{17, 122, 17}
\definecolor{cmblue}{rgb}{0.12156862745098039, 0.4666666666666667, 0.7058823529411765}
\definecolor{mygrey}{gray}{0.35}
\definecolor{myblue}{rgb}{0.2,0.2,0.8}
\definecolor{mygreen}{rgb}{0.2,0.8,0.5}
\definecolor{myzard}{cmyk}{0,0,0.05,0}
\definecolor{mywhite}{rgb}{1,1,1}
\definecolor{myred}{rgb}{1,0.,0.3}

\usepackage{graphicx}
\usepackage{dcolumn}
\usepackage{bm}
\usepackage[
colorlinks=true,
citecolor=myblue,
linkcolor=myred,
urlcolor=myblue,
]{hyperref}
\usepackage[mathlines]{lineno}
\usepackage{xcolor}
\usepackage{physics}
\usepackage{bbold}
\usepackage{mathtools}
\usepackage{tikz}
\usepackage{adjustbox}
\usepackage{tabularx}
\usepackage{algpseudocode}
\usepackage{algorithm}
\usepackage{cleveref}
\usetikzlibrary{quantikz}
\usepackage{soul}
\setstcolor{blue}

\newcommand{\Gn}{{G}_{\boldsymbol{n}}}
\newcommand{\Pn}{{P}_{\boldsymbol{n}}}

\begin{document}

\preprint{APS/123-QED}

\title{Noise-aware variational eigensolvers: a dissipative route for lattice gauge theories}

\author{J. Cobos}
\email{jesus.cobos@ehu.eus}
\affiliation{Department of Physical Chemistry, University of the Basque Country UPV/EHU, Box 644, 48080 Bilbao, Spain}
\affiliation{EHU Quantum Center, University of the Basque Country UPV/EHU, P.O. Box 644, 48080 Bilbao, Spain}

\author{D. F. Locher}
\email{d.locher@fz-juelich.de}
\affiliation{Institute for Quantum Information, RWTH Aachen University, 52056 Aachen, Germany}
\affiliation{Peter Grünberg Institute, Theoretical Nanoelectronics, Forschungszentrum Jülich, 52425 Jülich, Germany}

\author{A. Bermudez}
\email{bermudez.carballo@gmail.com}
\affiliation{Instituto de Física Teórica, UAM-CSIC, Universidad Autónoma de Madrid, Cantoblanco, 28049 Madrid, Spain}

\author{M. M\"uller}
\email{markus.mueller@fz-juelich.de}
\affiliation{Institute for Quantum Information, RWTH Aachen University, 52056 Aachen, Germany}
\affiliation{Peter Grünberg Institute, Theoretical Nanoelectronics, Forschungszentrum Jülich, 52425 Jülich, Germany}

\author{E. Rico} 
\email{enrique.rico.ortega@gmail.com}
\affiliation{Department of Physical Chemistry, University of the Basque Country UPV/EHU, Box 644, 48080 Bilbao, Spain}
\affiliation{EHU Quantum Center, University of the Basque Country UPV/EHU, P.O. Box 644, 48080 Bilbao, Spain}
\affiliation{Donostia International Physics Center, 20018 Donostia-San Sebastián, Spain}
\affiliation{IKERBASQUE, Basque Foundation for Science, Plaza Euskadi 5, 48009 Bilbao, Spain}

\date{\today}

\begin{abstract}
We propose a novel variational ansatz for the ground-state preparation of the $\mathbb{Z}_2$ lattice gauge theory (LGT) in quantum simulators. It combines dissipative and unitary operations in a completely deterministic scheme with a circuit depth that does not scale with the size of the considered lattice. We find that, with very few variational parameters, the ansatz can achieve $>\!99\%$ precision in energy in both the confined and deconfined phase of the $\mathbb{Z}_2$ LGT. We benchmark our proposal against the unitary Hamiltonian variational ansatz showing a  reduction in the required number of variational layers to achieve a target precision. After performing a finite-size scaling analysis, we show that our dissipative variational ansatz can predict accurate critical exponents without requiring a number of layers that scales with the system size, which is the standard situation for  unitary ans\"{a}tze. Furthermore, we investigate the performance of this variational eigensolver subject to circuit-level noise, determining variational error thresholds that fix the error rate below which it would be beneficial to increase the number of layers. In light of these quantities and for typical gate errors $p$ in current quantum processors, we provide a detailed assessment of the prospects of our scheme to explore the $\mathbb{Z}_2$ LGT on near-term devices.
\end{abstract}

\maketitle

\setcounter{tocdepth}{2}
\begingroup
\hypersetup{linkcolor=black}
\tableofcontents
\endgroup

\section{\bf Introduction}

Gauge theories are the mathematical formalism underlying the description of nature at its most fundamental level. Their origin is motivated by the study of phenomena in high-energy physics, where the Standard Model arises as an extremely successful gauge theory that has been tested with very high precision~\cite{schwartz_2013, cottingham_greenwood_2007}. They also appear in condensed-matter physics in the description of emergent phenomena, where the $\mathbb{Z}_2$ lattice gauge theory (LGT) arises as an effective model for high-$T_c$ superconductors~\cite{fradkin_2013}; and in quantum information, where some quantum error correcting codes are deeply related to LGTs~\cite{PhysRevB.108.045150}. Even if gauge theories are widely used in modern physics, most of them can not be solved exactly, and one has to resort to either perturbative expansions or numerical methods to extract quantitative predictions, especially in the case of non-Abelian local symmetries.

The perturbative approach for studying gauge theories has been extensively developed since their origin~\cite{schwartz_2013, cottingham_greenwood_2007}. However, it does not apply to phenomena that appear at large couplings, nor in the study of the infrared limit of some of these theories. The most notable example of this is the quest to quantitatively understand the mechanism driving the confinement of quarks~\cite{muta2010foundations, kogut_stephanov_2003}, which is a non-perturbative phenomenon that historically motivated the development of LGTs~\cite{rothe_book}. Despite the great success of LGTs, e.g.~\cite{doi:10.1126/science.1163233}, some important regimes remain inaccessible, as current quantum Monte Carlo techniques are afflicted by the sign problem at finite fermion densities and real-time dynamics~\cite{PhysRevLett.94.170201}.

\begin{figure*}[t]
    \includegraphics[width=0.9\linewidth]{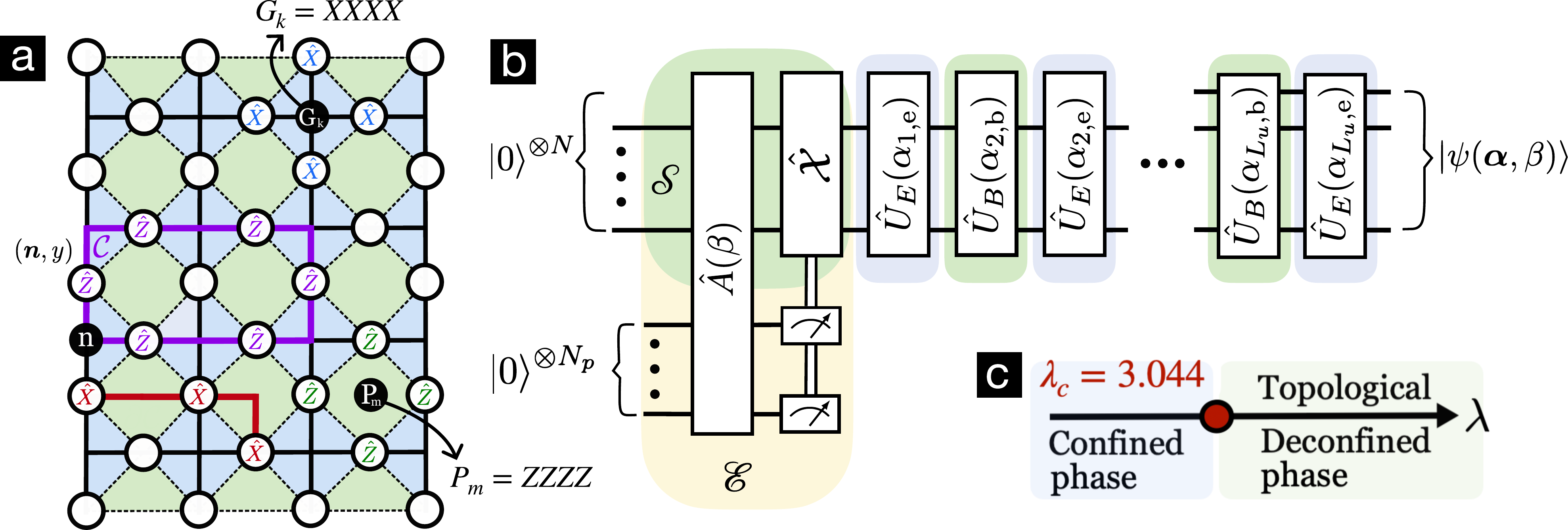}
    \caption{(a) Sketch of the pure $\mathbb{Z}_2$ LGT on a two-dimensional $d_x = 4$, $d_y = 5$ lattice with surface-code-like boundary conditions. White circles located on the links $(\boldsymbol{n}, i)$ of the lattice represent the gauge degrees of freedom, $\boldsymbol{n}$ labeling the vertices of the lattice and $i \in \{x, y\}$ indicating the orientation of a link. Matter fields would be placed on the vertices but are disregarded in the \emph{pure} $\mathbb{Z}_2$ LGT. Plaquette operators $\Pn$ and gauge transformation operators $\Gn$ are defined in Eqs.~\eqref{eq:plaquettes_def} and \eqref{eq:gauge_trans_def}, respectively. The red string in the figure represents a dual magnetization operator which acts as a non-local order parameter to label the different phases of the theory. In purple we sketch an example of a closed Wilson loop on the lattice. (b) The variational circuit is used to prepare the ground state of the $\mathbb{Z}_2$ LGT. In the first layer, a partial imaginary time evolution is implemented by a coupling ${A}(\beta)$ of an auxiliary system, which plays the role of an environment $\mathcal{E}$, to the physical qubits of the system $\mathcal{S}$, which encode the gauge degrees of freedom, and subsequently measuring the auxiliary degrees of freedom to induce effective dissipative dynamics, as shown in Eqs.~\eqref{eq:dissipative_steps_1}$-$\eqref{eq:dissipative_steps_4}. Depending on the ancilla measurement outcomes, a combination of Pauli-$X$ gates is applied to a subset $S$ of physical qubits, ${\mathcal{X}} = \bigotimes_{(\boldsymbol{n}, i) \in S} {X}_{(\boldsymbol{n}, i)}$, which results in the deterministic implementation of the partial imaginary time evolution. This dissipative step is followed by a sequence of unitary operations resembling the HVA~\eqref{eq:HVA} with $\ell_u$ layers, where ${U}_{E}(\alpha_{k, \mathrm{e}}) = {\rm e}^{{\rm i} \alpha_{k, \mathrm{e}} {H}_E}$ and ${U}_B(\alpha_{k, \mathrm{b}}) = {\rm e}^{{\rm i} \alpha_{k, \mathrm{b}} {H}_B}$. (c) Phase diagram of the $\mathbb{Z}_2$ LGT with confined and deconfined phases that are separated by a second-order Ising-like phase transition located at $\lambda_{\mathrm{c}} \approx 3.044$ \cite{critical_lambda}, as revealed by a non-local order parameter. Confinement is present for $\lambda < \lambda_c$, while deconfinement and topological order appear for $\lambda > \lambda_c$.}
    \label{fig:title}
\end{figure*}

The complexity of treating the non-Abelian gauge theories of the Standard Model led researchers to focus on simpler models, either in lower dimensions and/or simpler gauge groups, with the hope to identify the minimal ingredients that lead to paradigmatic phenomena, such as asymptotic freedom and confinement. Traditionally, the most widely studied models have been those in (1+1) spacetime dimensions, such as the Gross-Neveu model~\cite{PhysRevD.10.3235}, the Schwinger model for quantum electro-Dynamics ($\mathrm{QED}_2$)~\cite{PhysRev.128.2425, COLEMAN1975267}, or the t'Hooft model of quantum chromo-dynamics ($\mathrm{QCD}_2$) \cite{THOOFT1974461, Gross_1993}. In this work we consider the (2+1)-dimensional $\mathbb{Z}_2$ LGT~\cite{wegnerz2, kogutZ2, gauge_spin_models, Z2_transverse_fields, Z2_fermion_representation, hamiltonianZN}. This model contains both a confined phase and a deconfined phase that displays topological order and long-range entanglement~\cite{doi:10.1142/S0217979290000139, Wen_2013}. It is also closely related to the toric code \cite{KITAEV20032} and the surface code \cite{Dennis_2002, surface_codes_review, bravyi1998quantum,low_distance_surface_codes} in the context of quantum error correction (QEC) \cite{Gottesman_1998, Roffe_2019, Terhal_2015, Girvin_2023, gottesman2022opportunities}, where Gauss' law appears as an emergent super-selection rule in the ground-state manifold. The study of these simpler models is beneficial not only as a prelude to the more complex LGTs but also to develop a deeper understanding of other relevant phenomena, such as high-temperature superconductivity~\cite{PhysRevB.37.580} or frustrated magnetism~\cite{doi:10.1063/1.1665530}. Additionally, they can provide a more amenable arena to develop and benchmark new numerical simulations.

The search for alternative approaches for the simulation of LGTs, either classical or quantum, has been a very active area of research in recent years. In the classical simulation research line, tensor networks \cite{Banuls2020, Carmen_Banuls_2020} are a very promising way to avoid the sign problem, although they are only particularly efficient in situations with a limited amount of entanglement in the ground state and low-lying excitations. This renders them very effective for the study of static properties of LGTs, especially in low dimensions but limits their accuracy in predictions of real-time evolutions in principle \cite{Calabrese_2005}, due to the infamous entanglement barrier. Also, the tensor contraction operation that is extensively used in numerical algorithms is computationally expensive in high spatial dimensions. 
 
Quantum simulators (QSs)~\cite{Feynman_1982,Cirac2012,Bloch2012,Blatt2012,PRXQuantum.2.017003} offer a promising route for the study of unexplored regimes in LGTs~\cite{Banuls2020, PRXQuantum.4.027001, dimeglio2023quantum}. QSs are many-body systems that can be manipulated while maintaining their quantum properties and can be controlled to emulate the behavior of a target model under study. There are currently several platforms that are suitable to perform quantum simulations: cold atoms~\cite{doi:10.1126/science.aal3837, Schafer2020}, trapped ions~\cite{Dumitrescu2022, Zhang2017}, and either photonic~\cite{Wang2020, Aspuru-Guzik2012} or superconducting circuits~\cite{Houck2012, Dborin_2022, doi:10.1126/science.abi8794}, among others. Two different approaches are mainly considered: in analog quantum simulations, there exists a direct correspondence between the Hamiltonian describing the QS and the model that one wants to simulate. Some proposals of analog QSs for gauge theories can be found in Refs.~\cite{Banuls2020, doi:10.1098/rsta.2021.0064, PhysRevResearch.2.023015} and references therein. The other approach is a digital quantum simulation, where unitary operations are typically chosen from a universal set of quantum gates~\cite{nielsen00} and then composed to perform the simulation. The literature on digital simulation schemes for gauge theories is extensive~\cite{burrelo, mildenberger2022probing, Kalinowski_2023, PRXQuantum.3.020324}, see also references in \cite{Banuls2020, doi:10.1098/rsta.2021.0064, PRXQuantum.4.027001, dimeglio2023quantum}. During the current noisy intermediate-scale quantum (NISQ) era~\cite{Preskill_2018}, in which the quantum devices available are still limited by the presence of noise, it is expected that analog simulators will scale at a faster rate, due to their higher resilience to noise~\cite{trivedi2022quantum, Flannigan_2022}. However, the generality and eventual fault-tolerance of the digital approach may make it more useful in the long term. It may be the case that noise mitigation techniques \cite{PhysRevLett.119.180509, PhysRevX.8.031027, ferracin2022efficiently, Kim2023, majumdar2023best} enable digital QSs to get relevant results, even before fault-tolerant devices are available. Both approaches are certainly worth pursuing and each of them can benefit from the other~\cite{PRXQuantum.2.020328, Arrazola2016, PhysRevApplied.19.064086, garciadeandoin2023digitalanalog}. 

In this work, we present a dissipative variational quantum eigensolver~\cite{vqe_og, McClean_2016, Kandala2017, PhysRevX.7.021050, PhysRevX.8.031022, Yuan2019theoryofvariational, Kokail2019, Nakanishi_2019} for studying the properties of the $\mathbb{Z}_2$ LGT ground-state in quantum simulators. VQEs in general are a class of quantum-classical algorithms formulated in terms of a specific ansatz for the ground state of a quantum Hamiltonian, depending on a set of variational parameters, together with a procedure for preparing it in a quantum device. The ground state of the considered model is then approximated by finding the value of the variational parameters for which the energy of the resulting state is minimal. This is achieved through a feedback loop where the quantum device provides the value of the energy as the cost function, which is used by a classical optimization algorithm to find the optimal variational parameters. Once the optimal values of the variational parameters are determined, the ground-state preparation part is repeated on the QS as many times as one desires to measure relevant observables. Even though these algorithms are heuristic and do not provide guarantees of success, as they depend on the adequacy of the variational family of states, they are well-suited for current NISQ devices with short circuit depths~\cite{hubbard_vqe, cristian_vqe, heisemberg_vqe, hartree_fock_vqe, doi:10.1021/acsnano.5b01651, PhysRevA.95.020501, PhysRevX.8.011021}.

Other examples of widely used variational algorithms are the quantum approximate optimization algorithm~\cite{farhi2014quantum, blekos2023review, 8939749}, the Hamiltonian variational ansatz (HVA)~\cite{hva, hva_no_barren}, or the unitary coupled cluster~\cite{vqe_og, Yung2014, romero2018strategies, Anand_2022}. The first of them is most widely used in the context of optimization, while the others are intended for molecular and many-body simulations, respectively. These ans\"atze are known to have fundamental limits on their expressibility due to their unitary nature \cite{D0CP01707H, optimal_hva, 10.1063/1.5133059}. For example, the HVA approximates an optimal adiabatic evolution from an initial reference state to the desired ground state, but adiabaticity is known to break down across phase transitions \cite{optimal_hva, binkowski2023elementary}. This implies that different HVA ansätze must be used in each of the phases of the model under consideration, limiting their performance around phase transitions as one scales to larger system sizes. In this regard, the amount of entanglement in unitary ansätze  with a product reference state is generally limited by a Lieb-Robinson light-cone-like spreading, which sets limits on the scaling of the minimal number of variational layers that is required as the system size grows. This is of special importance in the vicinity of critical points, but also when considering topologically-ordered phases with long-range entanglement, such as the $\mathbb{Z}_2$ LTG. We note that, historically, considerable progress in variational methods for many-body quantum systems has been achieved by going beyond this unitary paradigm. Notable examples are the Gutzwiller~\cite{PhysRevLett.10.159, PhysRev.134.A923, PhysRev.137.A1726} and Jastrow-Gutzwiller~\cite{PhysRev.98.1479,doi:10.1143/JPSJ.59.3669,PhysRevLett.94.026406} ans\"atze for strongly-correlated electrons, the Jastrow-Marshall~\cite{PhysRevLett.60.2531,Horsch1988,PhysRevB.40.11437} ansatz for frustrated spin systems, or the resonating valence-bond ansatz for both frustrated spin systems~\cite{doi:10.1073/pnas.60.1.59,PhysRevLett.61.365} and high-temperature superconductivity~\cite{doi:10.1126/science.235.4793.1196,PhysRevB.38.931}. These ans\"atze can be understood as the normalized action of a non-unitary operator on a reference state which, for a certain limiting value of their variational parameters, acts as a projector onto the ground state of well-understood limits of the model under study. In addition, other well-known ans\"atze, such as the algebraic Bethe ansatz~\cite{Bethe1931,Sklyanin1979} and matrix product states~\cite{cmp/1104249404,PhysRevLett.75.3537} can also be understood as the effect of a non-unitary operation, see Refs.~\cite{Sopena2022algebraicbethe} and~\cite{10.5555/2011832.2011833}. 

A clear example of such a non-unitary operation with limiting projecting behavior is that of the propagator in imaginary time and its long-time limit. The implementation of this class of ans\"atze has remained outside the context of VQEs until recently \cite{McArdle2019, 10.1063/1.5027484, Motta2020, PRXQuantum.3.040305, Zhang_2022, PhysRevA.105.022440, Amaro_2022, PhysRevResearch.5.023174, chan2023simulating, cubitt2023dissipative, gong2023improved} as, in general, they require the incorporation of dissipative techniques into the VQE toolbox, often requiring access to a programable open-system dynamics \cite{diehl2008quantumstates, Verstraete2009, Barreiro2011, PhysRevX.6.011022, Hubisz_2021, Caballar_2014, chifang2023quantum} or, otherwise, ending up in probabilistic post-selected schemes that scale poorly with system size. It is worth mentioning that it is possible to approximate imaginary time evolution using unitary operations \cite{McArdle2019, Motta2020} but one then faces either a complex optimization problem or must be able to implement highly non-local operators. From a broader perspective, the actual utility of VQEs, either unitary or non-unitary, is mainly limited by the noise present in NISQ devices~\cite{barren_plateau_noise}. It is thus important to include noise in current assessments of VQEs, assessing the possible advantages by increasing the circuit depth. This will teach us important lessons about the realistic prospects of NISQ VQEs.

The dissipative VQE  proposed in this work addresses the limitations of the unitary HVA and improves on the required circuit depth to accurately capture the confinement-deconfinement transition of this LGT. This is achieved by introducing a non-unitary operation together with a deterministic implementation scheme with a circuit depth that does not scale with lattice size. We will show that this reduction in depth is essential to achieve reasonable results when performing the variational algorithm in the presence of circuit-level noise.

The article is structured as follows: in Sec.~\ref{sec:Z2}, we briefly review the pure $\mathbb{Z}_2$ LGT  and define our notation. We extend this short review and present some observables used to benchmark the variational ansatz in Appendix \ref{app:Z2}. In Sec.~\ref{sec:ansatz}, we present our dissipative VQE and its implementation in the circuit model. We analyze the performance of the dissipative variational ansatz (DVA) in the absence of noise in Sec.~\ref{sec:noiseless_results}. Here, the results of a state vector simulation of the ground-state preparation process are presented and compared with exact diagonalization. We also discuss how the DVA accounts for relevant observables, and perform a finite-size scaling analysis to show the accuracy of the ansatz around the critical region. These results showcase the superior performance of our non-unitary DVA in comparison to the unitary HVA. In Sec.~\ref{sec:results_noise}, we investigate the effect of circuit-level noise in both the DVA and HVA. We determine the variational error thresholds $p_\ell$ for an $\ell$-layer ansatz, which determines the level of physical error rates $p<p_\ell$ that must be attained such that a deeper ansatz with more than $\ell$ layers becomes beneficial in the presence of gate errors. We show that, under realistic noise conditions, our non-unitary DVA outperforms the unitary HVA. We also discuss a post-selection scheme to detect certain errors and improve further the performance of our DVA.

\section{\bf The $\mathbb{Z}_2$ lattice gauge theory (LGT)} \label{sec:Z2}

The pure $\mathbb{Z}_2$ LGT \cite{wegnerz2, kogutZ2, gauge_spin_models, hamiltonianZN} is one of the better-suited models for the first implementations of LGT QSs on NISQ devices. Despite its simple structure, this LGT  is not trivial at all: it contains two distinct phases connected by a phase transition that is not characterized by any local order parameter \cite{wegnerz2, elitzur}. One of these phases is confining, while the other one has long-range entanglement and topological order, as indicated in Fig.~\ref{fig:title}(c). The long-range entanglement introduces a lower bound on the minimal depth $\ell$ of any unitary circuit intended to prepare the exact ground state of this deconfined phase: it must scale with the size of the lattice $\ell \sim \mathcal{O}(d)$ \cite{lieb_robinson_topo, unitary_minimal_depth}. Besides this inherent complexity, understanding the properties and real-time dynamics of this gauge theory coupled with matter fields at finite densities still poses many open questions that would benefit from a QS. 

Let us start with a brief review of the $\mathbb{Z}_2$ LGT to introduce the main concepts and set our notation. In Appendix~\ref{app:Z2}, we extend this material and present some observables used as a benchmark for our VQE proposal. We consider the Hamiltonian formulation~\cite{PhysRevD.11.395} of the pure $\mathbb{Z}_2$ gauge theory on a two-dimensional spatial lattice. As illustrated in Fig.~\ref{fig:title}(a), the gauge degrees of freedom are located on the links of the lattice, labeled $(\boldsymbol{n}, i)$. The vectors $\boldsymbol{n} \in \mathbb{Z}_{d_x} \!\! \times \mathbb{Z}_{d_y}$ denote the vertices of the lattice and $i \in \{x, y\}$ indicates the axis of the unit vector $\{{x}, {y}\}$ connecting a vertex to a neighbouring one. The values $d_x$ and $d_y$ refer to the number of gauge links in the boundary of the lattice. We consider only square lattices with surface-code-like boundary conditions~\cite{low_distance_surface_codes}, as shown in Fig.~\ref{fig:title}(a), and thus set $d_x = d_y = d$ as the lateral distance of the square lattice $L=d$, which is referred to as the code distance in the context of QEC. Each of the $N = d^2 + (d-1)^2$ gauge fields is a two-level quantum system with a local Hilbert space $\mathcal{H}_{(\boldsymbol{n}, i)}=\mathbb{C}^2$ spanned by two basis states $\left\{ \ket{0}, \ket{1} \right\}$, which encode the elements of the $\mathbb{Z}_2$ gauge field. In the quantum simulation of this model, each of the local gauge degrees of freedom is represented by a single qubit. The full Hilbert space of the complete lattice is the tensor product of local Hilbert spaces $\mathcal{H} = \bigotimes_{\boldsymbol{n}, i} \mathcal{H}_{(\boldsymbol{n}, i)}$, and the physical subspace will be a certain super-selection sector defined below. The Kogut-Susskind Hamiltonian \cite{PhysRevD.11.395} for the pure $\mathbb{Z}_2$ LGT can be defined in analogy with the QED Hamiltonian~\cite{rothe_book} as 
\begin{equation}
    {H} = -{H}_E - \lambda {H}_B.
    \label{eq:hamiltonian}
\end{equation}
The analog of the electric energy term is 
\begin{equation}
\label{eq:elec_term}
    {H}_E = \sum_{\boldsymbol{n}, i} {X}_{(\boldsymbol{n}, i)},
\end{equation}
where ${X}_{(\boldsymbol{n}, i)}$ plays the role of the local electric-field operator and corresponds to the Pauli-$X$ matrix in the tensor-product Hilbert space. Since ${X}_{(\boldsymbol{n}, i)}^2=\mathbb{1}$, the electric-field energy is not quadratic but linear in the electric field, which is a peculiarity of the $\mathbb{Z}_2$ gauge theory in comparison to QED. This electric term competes with the magnetic-flux energy, which is proportional to 
\begin{equation}
\label{eq:mag_term}
    {H}_B = \sum_{\boldsymbol{n}} {P}_{\boldsymbol{n}} .
\end{equation}
The operators $\Pn$ are the Wilson plaquette operators
\begin{equation}
\label{eq:plaquettes_def}
    {P}_{\boldsymbol{n}}=  {Z}_{(\boldsymbol{n}, x)} {Z}_{(\boldsymbol{n} + {x}, y)} {Z}_{(\boldsymbol{n} + {y}, x)} {Z}_{(\boldsymbol{n}, y)} ,
\end{equation}
which correspond to the smallest gauge-invariant Wilson loops that can be defined on the lattice (green squares in Fig.~\ref{fig:title}(a)). They play the role of the magnetic flux piercing each of the  $N_p = d(d - 1)$  plaquettes of the lattice. We note that the plaquette operators on the top and bottom boundaries are tensor products of Pauli-$Z$ matrices acting only on three links. These are what we call surface-code-like boundary conditions in analogy to what can be found in Ref.~\cite{bravyi1998quantum,low_distance_surface_codes}. Using these boundary conditions is interesting because a degeneracy of the ground state will emerge for $\lambda \to \infty$, even for planar lattices, which is a consequence of the non-trivial homology of the model. The $\mathbb{Z}_2$ gauge transformations are generated by the so-called vertex operators
\begin{equation}
    {G}_{\boldsymbol{n}} = {X}_{(\boldsymbol{n}, -x)} {X}_{(\boldsymbol{n}, -y)} {X}_{(\boldsymbol{n}, x)} {X}_{(\boldsymbol{n}, y)} ,
    \label{eq:gauge_trans_def}
\end{equation}
which flip the basis states of every link connected to site $\boldsymbol{n}$. Gauge invariance implies that every local vertex operator commutes with the Hamiltonian~\eqref{eq:hamiltonian}
\begin{equation}
    \left[ {H}, {G}_{\boldsymbol{n}} \right] = 0, \hspace{0.5cm} \forall \; \boldsymbol{n}\in\mathbb{Z}_{d}\times\mathbb{Z}_{d} .
    \label{eq:gauge_invariance}
\end{equation}
This commutation relation indicates that the eigenvalues of the vertex operators are constants of motion. The Hermitian and involutory nature of the vertex operators ${G}_{\boldsymbol{n}}^2 = \mathbb{1}$ restricts their spectrum to be $\pm 1$. These eigenvalues are related to an analog of Gauss' law in the $\mathbb{Z}_2$ gauge theory~\cite{gauge_spin_models}, as they correspond to the presence ($-1$) or absence ($+1$) of a background $\mathbb{Z}_2$ charge at the corresponding lattice site $\boldsymbol{n}$. These conserved static charges serve to determine the different super-selection sectors of the gauge theory. Since we are considering the pure $\mathbb{Z}_2$ LGT, we define the {physical} Hilbert space as a subspace with vanishing static charges via the constraints
\begin{equation}
    {G}_{\boldsymbol{n}} \ket{\psi} = \ket{\psi}, \hspace{0.5cm} \forall \; \boldsymbol{n}\in\mathbb{Z}_d\times\mathbb{Z}_d.
    \label{eq:physical_states}
\end{equation}

The microscopic coupling $\lambda$ in Eq.~\eqref{eq:hamiltonian} is a real parameter, the value of which shall determine the specific zero-temperature phase of the $\mathbb{Z}_2$ LGT. As discussed in more detail in Appendix~\ref{app:Z2}, when this coupling lies below a certain critical value $\lambda<\lambda_{c} \approx 3.044$~\cite{critical_lambda}, the ground-state is a short-range entangled state which, for $\lambda \to 0$, reduces to a trivial product state
\begin{equation}
    \label{eq:electric_gs}
    \ket{\Omega_E} = \bigotimes_{\boldsymbol{n}, i} \ket{+_{(\boldsymbol{n},i)}},
\end{equation}
In this regime, one says that the gauge field is confining. On the contrary, the magnetic coupling dominates for $\lambda > \lambda_{c}$, and one observes long-range entanglement, a topological degeneracy, and no confinement in the ground state. The lowest eigenstate of the magnetic term is obtained by applying a projector onto the common $+1$-eigenspace of the plaquette operators to the previous state
\begin{equation}
    \label{eq:magnetic_gs}
    \ket{\Omega_B} = \prod_{\boldsymbol{n}} \frac{1}{\sqrt{2}}\!\left(\mathbb{1} + {P}_{\boldsymbol{n}} \right)\ket{\Omega_E} .
\end{equation}
Fulfilling also Gauss' law (Eq.~\eqref{eq:physical_states}), the state $\ket{\Omega_B}$ can thus be interpreted as a stabilizer state in the code space of the surface code. We will come back to this connection in a later section. Note that $\ket{\Omega_B}$ is just one state in the lowest-eigenvalue subspace of ${H}_B$. The dimension of this subspace depends on the topology of the surface in which the lattice is embedded, or on the nature of the boundary conditions for a planar lattice. The remaining states in this subspace can be generated by applying strings of ${Z}_{(\boldsymbol{n}, i)}$-operators along non-contractible loops of the lattice \cite{KITAEV20032}. In Appendix~\ref{app:Z2} we extend the discussion on the physical Hilbert space of the $\mathbb{Z}_2$ LGT.

The simulation of the two-dimensional $\mathbb{Z}_2$ LGT  is an ideal benchmark for digital QSs, as it presents many of the challenges that will be met in other higher-dimensional non-Abelian gauge theories -- but these are incarnated at their simplest possible form, as the gauge degrees of freedom are encoded in single qubits, and the locality of the Hamiltonian only requires local gate connectivities. In addition, various observables can be inferred from standard projective Pauli-basis measurements,  giving access to characteristic non-trivial effects in the LGT. These observables serve as markers of the confined-deconfined phase transition and allow one to test topological order in the deconfined phase (see Appendix~\ref{app:Z2}). In analog QSs, the implementation of the plaquette terms \eqref{eq:plaquettes_def} is a bottleneck,  as it requires access to four-body interactions, which are typically perturbatively small~\cite{Banuls2020, halimeh2023spin, Katz2023, Dai2017}. Alternatively, one can use the fact that the deconfining effect of the plaquette terms can sometimes arise from energetic considerations when including matter~\cite{PhysRevX.10.041007} to perform simulations without the need to implement 4-body interactions \cite{bazavan2023synthetic,domanti2023floquetrydberg}. This caveat is overcome in digital QSs~\cite{RevModPhys.82.2313, Bruzewicz_2019, Krantz_2019, Slussarenko_2019}, as higher-weight terms can be obtained by nested entangling gates. On the other hand, the number of Trotter steps required for the full temporal evolution with magnetic and electric terms in digital QSs typically requires large circuit depths~\cite{gate_count_ltg}, which can be limited by the accumulation of gate errors.

The problem of ground-state preparation for the $\mathbb{Z}_2$ LGT in digital simulators has been tackled before using VQEs \cite{burrelo}. However, the theoretical studies of VQEs do not typically incorporate the fact that the gates used to build each layer will inevitably be faulty and the errors will propagate to other layers, proliferating due to the underlying non-fault-tolerant construction of the circuits. In the following section, we present our proposal for our non-unitary DVA  for the $\mathbb{Z}_2$ LGT and show that it does improve upon the standard HVA approach \cite{burrelo} both in the noiseless and noisy regimes. Our study provides a solid foundation for the development of scalable VQEs for lattice gauge theories in near-term quantum devices.

\section{\bf Non-unitary ans\"atze for gauge theories} \label{sec:ansatz}

\subsection{Dissipative variational quantum eigensolvers~(VQEs)}

In this section, we introduce a non-unitary variational ansatz to approximate the ground state of the $\mathbb{Z}_2$ LGT. We provide an algorithm for the deterministic preparation of this quantum state in digital quantum computers so that it can be used as a dissipative VQE. VQEs are a family of hybrid quantum-classical algorithms whose goal is to find the eigenstates of target Hamiltonians. The basic principle underlying these algorithms is simple: any state $\ket{\psi(\vb*{\alpha})}$ of a quantum system different from the ground-state(s) $\ket{E_0}$, and defined in terms of a set of variational parameters $\boldsymbol{\alpha}$, will have mean energy that is larger than the latter $E_0$ \cite{McClean_2016}, such that
\begin{equation}
\label{eq:var_principle}
   E(\vb*{\alpha})= \bra{\psi(\vb*{\alpha})} {H} \ket{\psi(\vb*{\alpha})} > E_0.
\end{equation}
In traditional VQEs, one tries to approximate this ground state by preparing the actual ansatz on a physical device, which is accomplished by acting with a series of unitary operations that depend on the set of variational parameters on an easy-to-prepare reference state $\ket{\psi_0}$. Each of these unitaries is built from a set of gates that depend on a certain parameter that can be experimentally manipulated \cite{doi:10.1080/00107514.2011.603578, Krantz_2019, Lim2014}. The resulting  state is
\begin{equation}
\label{unitary_vqe}
    \ket{\psi_{\rm u}(\vb*{\alpha})} = {U}_{\vb*{\alpha}_{\ell_{u}}} \,{U}_{\vb*{\alpha}_{\ell_{u}-1}}\, \cdots \, {U}_{\vb*{\alpha}_{1}} \ket{\psi_0} 
\end{equation}
where $\vb*{\alpha}=(\vb*{\alpha}_{\ell_u},\vb*{\alpha}_{\ell_u-1}, \cdots,\vb*{\alpha}_{1})$ is a vector of the variational parameters in all the $\ell_{u}$ unitary layers.

Note that the product of unitary operations in Eq.~\eqref{unitary_vqe} is not the most general operation allowed by quantum mechanics. It is indeed possible to define a more general version of the VQE using $\ell_d$ dissipative layers
\begin{equation}
\label{eq:mixed_state_var}
    {{\rho}(\vb*{\beta})} = {\mathcal{E}}_{\beta_{\ell_d}} \circ {\mathcal{E}}_{\beta_{\ell_d-1}}\,\circ\, \dots \, \circ \,{\mathcal{E}}_{1}\big(\ket{\psi_0}\!\!\bra{\psi_0}\big),
\end{equation}
each of which corresponds to a quantum channel described by a completely positive trace-preserving (CPTP) map \cite{nielsen00}. These channels can be mathematically expressed by the so-called Kraus decomposition
\begin{equation}
\label{eq:channel_kraus}
    {\mathcal{E}}_{{\beta}_j}\big({\rho_0}\big)=\sum_{n=1}^\kappa {K}^{\phantom{\dagger}}_{n,\beta_j}\rho_0\, {K}^\dagger_{n, \beta_j}\!:\hspace{1ex}\sum_{n=1}^\kappa {K}^\dagger_{n,\beta_j}{K}^{\phantom{\dagger}}_{n,\beta_j}=\mathbb{1},
\end{equation}
where $\kappa$ is the so-called Kraus rank. For $\kappa=1$, we have a single Kraus operator that must be unitary, such that the variational family of states~\eqref{eq:mixed_state_var} reduces to that of the unitary VQE in Eq.~\eqref{unitary_vqe} with $\vb*{\beta}=\vb*{\alpha}$. For larger ranks, the quantum channels~\eqref{eq:channel_kraus} are typically used to describe the dissipative dynamics that result from the coupling of a quantum system to a larger environment. 
 
As discussed in Ref.~\cite{nielsen00}, given a certain Kraus rank $\kappa$, one can model the effect of any CPTP map with a unitary coupling ${U}_{\vb*{\beta}_j}$ between the system and an ancillary $\kappa$-level system $\{\ket{e_n}\}_{n=1}^\kappa$, after tracing over the ancillary system, such that $K_{n,\beta_{j}}={\rm Tr}_{\rm aux}\{{U}_{\beta_{j}}(\ket{\psi_0}\!\!\bra{\psi_0}\otimes\ket{e_n}\!\!\bra{e_n}){U}^\dagger_{\beta_{j}}\}$. In general, after this partial trace, the evolution of the reduced system will not preserve the purity of the initial reference state $\ket{\psi_0}$. We note that the above variational family~\eqref{eq:mixed_state_var} with purity non-preserving CPTP maps departs from a variational ground-state preparation algorithm, where one is interested only in pure states~\eqref{eq:var_principle}. In that respect, provided one could measure the auxiliary system and post-select on a specific result, e.g.  $n=1$, the density matrix~\eqref{eq:mixed_state_var} would reduce to a pure state   
\begin{equation}
\label{non_unitary_vqe}
    \ket{\psi_{\rm d}(\vb*{\beta})} = \frac{1}{\mathcal{N(\vb*{\beta})}}{K}_{1,\beta_{\ell_d}} \,{K}_{1,\beta_{\ell_d-1}}\, \cdots \, {K}_{1,\beta_{\ell_1}} \ket{\psi_0}, 
\end{equation}
where  the normalization constant $\mathcal{N}(\vb*{\beta})$ can be interpreted as the post-selected probability amplitude. This ancillary method corresponds to a post-selected positive operator-valued measure \cite{watrous_2018}, where the resulting individual operators are neither unitary nor orthogonal projectors. Indeed, one could combine the unitary and post-selected dissipative dynamics to define the ansatz
\begin{equation}
\label{dissipative_vqe}
    \ket{\psi_{\rm d}(\vb*{\alpha},\vb*{\beta})} = \frac{1}{\mathcal{N(\vb*{\beta})}}\prod_{j=1}^{\ell_u} {U}_{\alpha_j}\prod_{k=1}^{\ell_d}{K}_{1,\vb*{\beta}_k}  \ket{\psi_0}, 
\end{equation}
where the products are decreasing in the layer index. One can even intertwine the unitary and non-unitary operations, leading to a more general pure state. We call this ansatz the {\it dissipative variational ansatz} (DVA), and it will be the subject of our work. 

As advanced in the introduction, the motivation to consider this type of VQEs is that there exists a variety of useful ans\"atze in quantum many-body systems that cannot be described by a unitary operation acting on a reference state. Well-known examples appear in strongly correlated electrons under the name of the Gutzwiller \cite{PhysRevLett.10.159} and Jastrow \cite{PhysRev.98.1479} ans\"atze. For instance, the Gutzwiller ansatz would fall into the class of $\ell_u=0, \ell_d=1$, with $K_{1,\beta}$ being the so-called Gutzwiller operator that penalizes double occupancies of fermions. The goal of the current work is to explore the performance of this type of dissipative VQEs~\eqref{dissipative_vqe} in the context of the $\mathbb{Z}_2$ LGTs and analyze their performance when the unitary and dissipative operations are both affected by noise in NISQ devices.  

Even if, in principle, such a non-unitary ansatz~\eqref{dissipative_vqe} can be an extremely powerful calculation tool, we need to address a crucial point: the post-selection procedure. As the number of dissipative layers $\ell_d$ increases, the probability of obtaining the desired outcome from the measurement of the ancillary system will become vanishingly small \cite{leadbeater2023nonunitary}, making the experimental procedure rather inefficient. This is more daunting if one considers that, eventually, one is interested in the thermodynamic limit of a certain quantum many-body model, where the post-selection probability of even a single dissipative layer can also become negligible in the limit of large lattice sizes. One of the key results of our work is to show that, for a certain DVA of the $\mathbb{Z}_2$ LGT, we can devise a feed-forward operation depending on the outcome of the ancillary-system measurement that allows us to prepare the ansatz deterministically for $\ell_d=1$, arbitrary $\ell_u$, and arbitrary system size~$d$. We provide the details below.

Before turning to that specific discussion, let us note that, in a VQE, the values of the variational parameters $(\vb*{\alpha},\vb*{\beta})$ are found by minimizing the expectation value of the energy $E(\vb*{\alpha}, \vb*{\beta})$~\eqref{eq:var_principle} of the resulting variational state. The energy minimization is performed using a quantum-classical feedback loop in which the NISQ device provides the expectation value to a classical optimizer that uses it as its cost function. Recently, the effectiveness of this process has been brought into question. First of all, its variational nature impedes the formulation of general complexity arguments. In addition, the strong non-convexity of the cost function, together with the presence of Barren plateaus \cite{barren_plateau_og, barren_plateaus_shallow, barren_plateau_noise}, complicate the energy minimization process. An interesting discussion about the complexity of VQE training for a particular problem can be found in Ref.~\cite{vqe_np_hard}. It is also worth mentioning that it is possible to formulate ans\"atze capable of avoiding Barren plateaus, at least under certain conditions and, importantly, under the idealized assumption that there are no errors. Apart from these difficulties, researchers have empirically found cases in which unitary VQEs can approximate ground-states of complex systems with good fidelity \cite{vqe_og, burrelo, hartree_fock_vqe, heisemberg_vqe, hubbard_vqe, cristian_vqe}. Additionally, the variational states resulting from VQEs can be used as reference states for more resource-demanding algorithms for ground-state preparation that have guarantees of success \cite{filters_cirac, filters_lin_tong}. In the following section, we will show how our dissipative VQE can achieve extremely precise ground-state energy estimates without encountering Barren plateaus, which we believe is due to the reduced number of layers that are required and the combination of both unitary and non-unitary operations.

\subsection{Dissipative VQE for the $\mathbb{Z}_2$ LGT} \label{subsec:dissipative_z2}

We propose to use a version of the DVA~\eqref{dissipative_vqe} for the $\mathbb{Z}_2$ LGT with a single non-unitary layer 
\begin{equation}
    \begin{split}
        \ket{\psi_{\rm d}(\vb*{\alpha}, {\beta})} & = 
        \prod_{j = 2}^{\ell_u} \left( {\rm e}^{{\rm i} \alpha_{j,{\rm e}} {H}_E} {\rm e}^{{\rm i} \alpha_{j,{\rm b}} {H}_B} \right) 
        {\rm e}^{{\rm i} \alpha_{1,{\rm e}} {H}_E}  {{\rm e}^{\beta {H}_B}}\frac{\ket{\Omega_E}}{{{\mathcal{N}(\beta)}}},
    \end{split}
    \label{eq:ansatz}
\end{equation}
where the reference state is the trivial product state~\eqref{eq:electric_gs}, the normalization constant is $\mathcal{N}(\beta)=(\cosh 2\beta)^{N_p/2}$, and we recall that $N_p=d(d - 1)$ is the number of plaquettes (see Fig.~\ref{fig:title}{(a)}). This ansatz, which is gauge invariant,  falls into the class of the DVA  introduced in Eq.~\eqref{dissipative_vqe}, where we assume a single dissipative layer $\ell_d=1$, and an arbitrary number of unitary layers $\ell_u$, each of which is composed by the Trotterized construction for the $\mathbb{Z}_2$ gauge theory ${H}=-{H}_E-\lambda{H}_B$. The first unitary Trotter operation only contains the electric-field term~\eqref{eq:elec_term}, whereas the remaining ones, $j\geq2$, contain both the electric and magnetic terms with different variational parameters $\{\alpha_{j,{\rm e}},\alpha_{j,{\rm m}}\}$. A different way of interpreting ansatz \eqref{eq:ansatz} is to consider the completely unitary HVA  
\begin{equation}
    \ket{\psi_{\rm u,e}(\vb*{\alpha}, \alpha_{1,{\rm m}})} =  \prod_{j = 1}^{\ell_u} {\rm e}^{{\rm i} \alpha_{j,{\rm e}} {H}_E} {\rm e}^{{\rm i} \alpha_{j,{\rm b}} {H}_B}  \ket{\Omega_E},
    \label{eq:HVA}
\end{equation}
in which the variational parameter of the first Trotter component is Wick rotated $\alpha_{1,{\rm m}}\mapsto-{\rm i}\beta$, such that $ \ket{\psi_{\rm u,e}(\vb*{\alpha},\alpha_{1,{\rm m}})}\mapsto \ket{\psi_{\rm d}(\vb*{\alpha},\beta)}$. In Sec.~\ref{sec:noiseless_results}, we demonstrate that the introduction of the non-unitary layer in our ansatz \eqref{eq:ansatz} enables the DVA to converge close to the true ground state energy with considerably fewer layers, enabling it to scale to larger lattices while maintaining the number of gates amenable for implementation. In the following, we provide theoretical arguments supporting these results. For this, we first present some limitations of the HVA, and then discuss how they can be overcome with the introduction of the non-unitary layer.

The unitary HVA consists of applying a series of $\ell_u$ Trotterized propagators on a reference state. This state corresponds to the ground state of one of the terms in the considered Hamiltonian and it is assumed to be efficiently prepared in the NISQ device. Instead of fixing the time of evolution of each Trotter term from a fixed adiabatic schedule and posterior discretization of time, they are left as independent variational parameters $\boldsymbol{\alpha}$. The use of the HVA for the $\mathbb{Z}_2$ LGT  has already been thoroughly investigated in the relevant work of L. Lumia et {\it al.}~\cite{burrelo}, showing that it can attain a considerable accuracy in terms of energy and state fidelity for the electric-field dominated phase. This accuracy rests on the fact that this Trotter-like construction can approximate an optimal adiabatic evolution, up to errors controlled by the number of Trotter-layers (variational parameters) that one wants to consider \cite{optimal_hva}. In the particular case of \eqref{eq:HVA}, it simulates an adiabatic evolution from $\lambda = 0$ up to a specific non-zero value $\lambda > 0$ considered in the minimization of the energy $E(\vb*{\alpha},\alpha_{1,{\rm m}})$. As mentioned in the introduction,  any HVA faces an important obstacle: adiabaticity breaks down when crossing phase transitions, which limits the usefulness of the HVA ansatz to a single phase for large systems. A possibility discussed in~\cite{burrelo} to go around this limitation is to change the reference state and the Trotter order, yielding  another ansatz 
\begin{equation}
    \ket{\psi_{\rm u, m}(\vb*{\alpha}, \alpha_{1,{\rm e}})} =  \prod_{j = 1}^{\ell_u} {\rm e}^{{\rm i} \alpha_{j,{\rm b}} {H}_B} {\rm e}^{{\rm i} \alpha_{j,{\rm e}} {H}_E} \ket{\Omega_B}.
    \label{eq:HVA_B}
\end{equation}
However, one must bear in mind that a full unitary HVA protocol for reference states with non-trivial entanglement content can dramatically increase the circuit complexity, as occurs for the deconfined state $\ket{\Omega_B}$~\eqref{eq:magnetic_gs}, which would require a number of unitary layers that scales with the size of the lattice $\ell_u \sim \mathcal{O}(d)$~\cite{lieb_robinson_topo, unitary_minimal_depth}. We note that, if one chooses to use ancilla-qubit measurements, one can instead prepare this reference state in constant depth \cite{topological_state_ions}, as is known also for  QEC with the surface code. 

Even in this situation, a shallow HVA~\eqref{eq:HVA_B} is now only accurate for $\lambda>\lambda_{c}$ and exhibits similar limitations around the critical region of the system as Eq.~\eqref{eq:HVA}. As we show in Sec.~\ref{sec:noiseless_results}, both HVAs fail at capturing accurately the most interesting region of the $\mathbb{Z}_2$ LGT: the confinement-deconfinement transition. Our proposal incorporates a variational dissipation into the ansatz itself, rather than in the preparation of a reference state, and can overcome these limitations providing a promising path towards exploring the criticality of LGTs in near-term devices. This is achieved through the introduction of a single non-unitary operation ${K}_{1,\beta}=e^{\beta {H}_B}$ in Eq.~\eqref{dissipative_vqe}, which is the Wick-rotated version of the first Trotter step of the magnetic term of the Hamiltonian. Expanding the non-unitary exponential in power series and using the involutory property of the plaquette operators ${P}_n^2 = \mathbb{1}$,  
the action of the non-unitary operator in the DVA
\begin{equation}
    \label{eq:mag_tem_ans}
    {K}_{1,\beta}=\prod_{\boldsymbol{n}} (\mathbb{1} \cosh\beta+ {P}_{\boldsymbol{n}}\sinh\beta) ,
\end{equation}
is indeed to partially project the state $\ket{\Omega_E}$ onto the magnetic ground-state $\ket{\Omega_B}$~\eqref{eq:magnetic_gs}.
In the limit
\begin{equation}
    \frac{{K}_{1,\beta}}{\mathcal{N}(\beta)} \ket{\Omega_E} \xrightarrow[\beta \to \infty]{} \ket{\Omega_B} = \prod_{\boldsymbol{n}}\left[\frac{1}{\sqrt{2}}(\mathbb{1}+{P}_{\boldsymbol{n}}) \right] \ket{\Omega_E},
\end{equation}
this operator completely projects out all the $-1$ eigenstates of the magnetic plaquettes. In the language of quantum information, this operator becomes a projector onto the common stabilizer subspace of the magnetic operators. On the other hand, $K_{1,\beta}\to\mathbb{1}$ as $\beta\to 0$, such that the DVA  interpolates between the electric~\eqref{eq:electric_gs} and magnetic~\eqref{eq:magnetic_gs} ground-states. For $\beta\in(0,\infty)$, the non-unitary operator ${K}_{1,\beta}$ is not an orthogonal projector, and aims at approximating the ground-state of the $\mathbb{Z}_2$ LGT for arbitrary values of $\lambda$ by allowing for a certain admixture of magnetic $\pi$-flux  plaquettes. We note that the non-unitary operator ${K}_{1, \beta}$ was first considered in the works \cite{cardy, PhysRevD.29.300,PhysRevD.32.1491}, where it was shown that it suffices to capture a second-order confinement-deconfinement phase transition. However, this approach overestimates the value of the critical coupling $\lambda_{\rm c} = 4$ and, more importantly, yields mean-field critical exponents $\beta=\nu=1/2$ \cite{goldenfeld}, which are known to differ from those of the $\mathbb{Z}_2$ LGT.

In this work, we show that, by enlarging the variational manifold with the unitary layers of Eq.~\eqref{eq:ansatz}, we can overcome all of these limitations and get a much more accurate DVA. In addition, we will also show in the next section that the full DVA, including the non-unitary layer, can be implemented deterministically using standard circuits of quantum computation with gates between the qubits encoding the gauge fields and ancillary qubits, which are measured projectively yielding classical information that is feed-forwarded to apply additional single-qubit gates on the system qubits.

\subsection{Deterministic circuit implementation}\label{sec:circuit_implementation}

In many near-term quantum processors, qubits are arranged in planar geometries which makes them suitable platforms to study LGTs in two-dimensional lattices. Superconducting transmons, for example, are placed on a chip in a fixed geometry, and tunable couplers enable the execution of two-qubit gates between nearest neighbors. Neutral atoms in optical tweezer arrays constitute another promising platform for digital quantum computation. Hundreds of atoms can be arranged in arbitrary geometries in two dimensions without defects \cite{Endres2016Atom, Barredo2016Anatom} and they can even be dynamically rearranged during a computation \cite{rydber_atoms_processor}. Temporarily exciting atoms in Rydberg states enable the execution of entangling gates between qubits that may reach beyond their nearest neighbors. The variational algorithm presented in this paper can be executed on a quantum processor which, besides access to single-qubit rotations and a two-qubit gate, has the following capabilities: {\it (i)} It can accommodate a 2-dimensional lattice of qubits with nearest-neighbour connectivity, either directly -- owing to its architectural design -- or through long-range interactions or shuttling \cite{rydber_atoms_processor, kaushal2020shuttling, moses2023race}. {\it (ii)} It can perform unitary operations on a subset of qubits conditioned on the result of a measurement. The latter capability is necessary to implement the dissipative operation deterministically. These requirements are in close correspondence to those needed to implement some of the most common topological quantum error correction codes, especially the different flavors of the surface code \cite{KITAEV20032, Dennis_2002, surface_codes_review, low_distance_surface_codes}. Topological states have already been prepared in experiments with superconducting qubits as well as in neutral atom and ion trap platforms \cite{topological_state_google, Semeghini2021probing, rydber_atoms_processor, topological_state_ions}. Moreover, multiple rounds of quantum error correction have been demonstrated in the surface code using superconducting qubits~\cite{surface_code_markus_wallraff, Zhao2022Realization, surface_code_google} and in the color code with trapped ions~\cite{ryananderson2021realization}.

\begin{figure}
    \centering
    \includegraphics[scale=0.7]{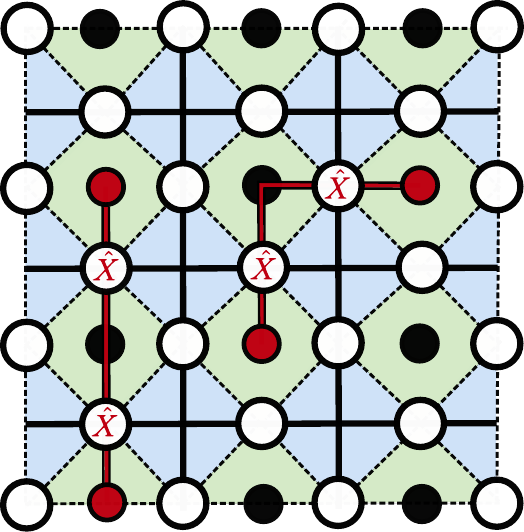}
    \caption{Configuration of physical qubits in a quantum processor for the investigation of the $\mathbb{Z}_2$ LGT. White circles located on the lattice links correspond to qubits encoding the gauge degrees of freedom. Black and red circles represent ancilla qubits located in the centers of the plaquettes (green squares). Ancilla qubits are required to implement the dissipative operation shown in Fig.~\ref{fig:non_unitary_implementation}. The red qubits represent an example of ancillae that have been projected onto the state $\ket{1}_a$ in the measurement. Such excited plaquettes must be removed by matching two of them and applying ${X}$-operators to all gauge qubits along that path. Instead of connecting two excited plaquettes, it is also possible to choose paths that end at the right or left boundary. Thus, it is possible to implement the dissipative operation even if an odd number of ancillae is measured in $\ket{1}_a$.}
    \label{fig:plaquettes_correction}
\end{figure}

The implementation of the non-unitary part of the ansatz in Eq.~\eqref{eq:ansatz} requires the introduction of ancilla qubits. The exact number of required ancillae depends on the qubit connectivity and their reusability after measurements. In a device with all-to-all connectivity and full reusability, a single ancilla qubit in principle suffices. However, the introduction of $N_p$ ancilla qubits is beneficial, since it provides an opportunity for parallelization and it can relax the connectivity requirements to nearest-neighbor interactions. In the following, we will assume access to a device in which qubits are distributed in a two-dimensional square lattice, as shown in Fig.~\ref{fig:plaquettes_correction}. The qubits representing the gauge degrees of freedom in the $\mathbb{Z}_2$ LGT are located on the links of a square lattice. Choosing this lattice to have twice the lattice constant as compared to the physical qubit lattice leaves qubits in the centers of the plaquettes that are used as ancillas and free qubits in the vertices.

\begin{figure}
  \centering
  \includegraphics[width=1\linewidth]{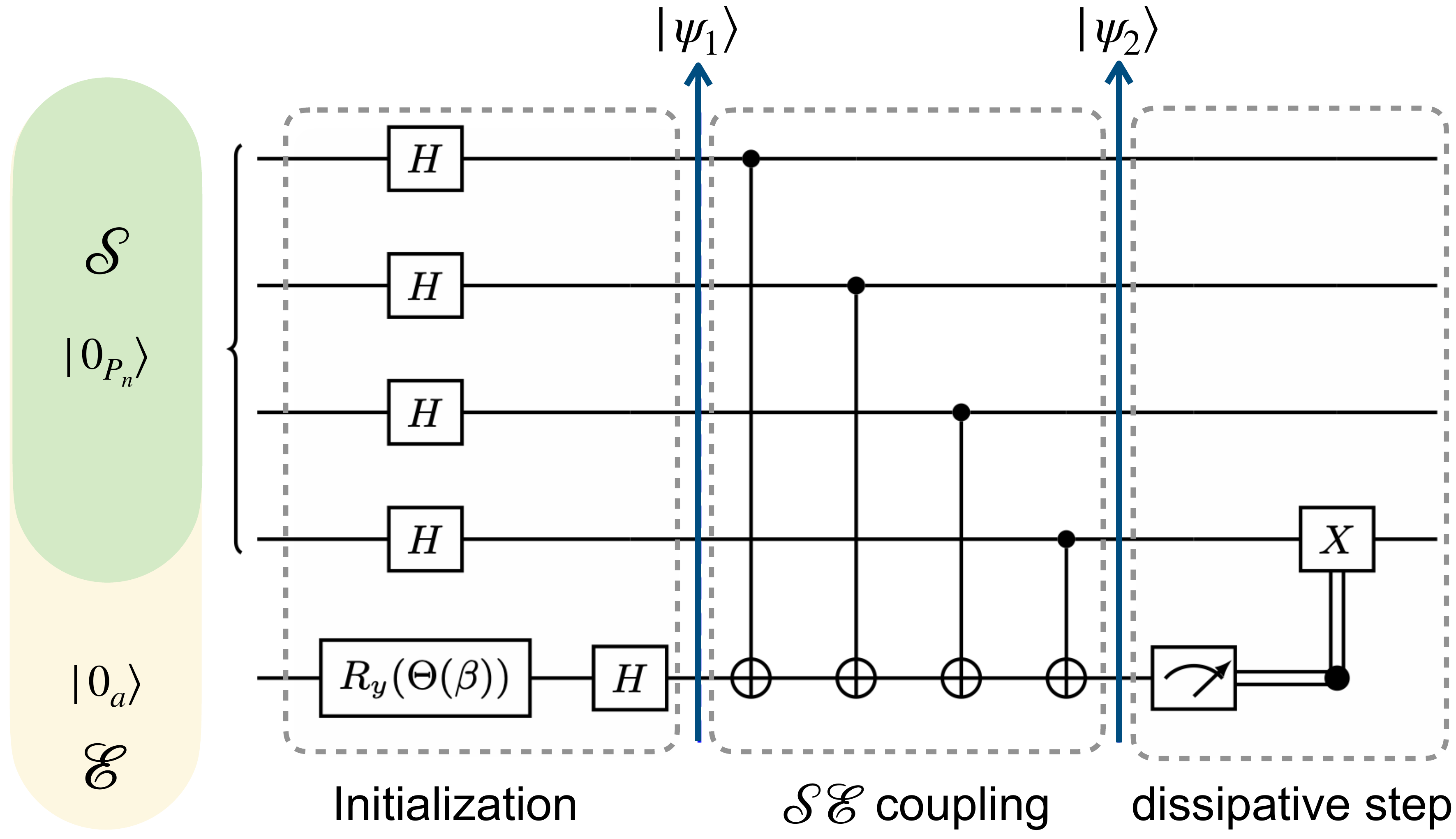}
  \caption{Circuit implementation of the non-unitary exponential of a single plaquette acting on the electric ground state: $(\cosh 2\beta)^{-1/2}\,{\rm e}^{\beta_1 \Pn} \ket{\Omega_E}_{P_{\boldsymbol{n}}}$. $\mathcal{S}$ is the set of qubits encoding the state of the plaquette qubits, while $\mathcal{E}$ hosts the ancilla qubit(s) that will play the role of an environment. The rotation angle acting over the later is $\Theta(\beta) = 2\tan^{-1} (\tanh \beta)$. The classically conditioned gate ${X}_{(\boldsymbol{n}, i)}$ is only applied if the ancilla qubit $\mathcal{E}$ is projected onto the state $\ket{1}_a$ in the measurement. When implementing the full non-unitary exponential~\eqref{eq:ansatz}, the classically conditioned operation depends on the set of all ancilla measurements, as discussed in the main text.}
  \label{fig:non_unitary_implementation}
\end{figure}

We propose the following scheme for the preparation of the DVA \eqref{eq:ansatz}. Since the plaquette operators $\Pn$ commute mutually, the non-unitary propagator in Eq.~\eqref{eq:ansatz} can be exactly factorized as the product of exponentials of single plaquettes (Eq.~\eqref{eq:mag_tem_ans}). Figure~\ref{fig:non_unitary_implementation} shows the quantum circuit implementing the exponential $(\cosh 2\beta)^{-1/2} \, {\rm e}^{\beta {P}_{\boldsymbol{n}}}$ acting on the reference state $\ket{\Omega_E}_{P_{\boldsymbol{n}}}$. Step by step, the action of this circuit is as follows: the first set of single-qubit operations prepares the physical qubits in the state $\ket{\Omega_E}$ and rotates the ancilla by an angle $\Theta(\beta) = 2 \tan^{-1} (\tanh \beta)$ around the $y$-axis on the Bloch sphere $R_y(\Theta)={\rm exp}\{-{\rm i}\Theta Y_a/2\}$, where $Y_a={\rm i} X_aZ_a$. This rotation encodes each of the constants in front of the identity and the plaquette operator in Eq.~\eqref{eq:mag_tem_ans}.
The state $\ket{\psi_1}$, indicated in Fig.~\ref{fig:non_unitary_implementation}, thus reads
\begin{equation}
         \ket{\psi_1} = \left( \frac{\ket{+}_a + \tanh \beta \ket{-}_a}{\sqrt{1 + \tanh^2 \beta}} \right) \otimes\ket{\Omega_E}_{P_{\boldsymbol{n}}}.
         \label{eq:dissipative_steps_1}
\end{equation}
After that, the train of CNOTs introduces the $\Pn$ operator in front of the state $\ket{-}_a$.
This is the case because each CNOT flips the sign in front of $\ket{-}_a$ if the corresponding physical qubit is in the state $\ket{1}_{(\boldsymbol{n}, i)}$. This introduces a product of ${Z}_{(\boldsymbol{n}, i)}$ in front of $\ket{-}_a$ which results in the operator $\Pn$. We obtain
\begin{equation}
    \ket{\psi_2} = \frac{\ket{+}_a \ket{\Omega_E}_{P_{\boldsymbol{n}}} + \tanh \beta \ket{-}_a \Pn \ket{\Omega_E}_{P_{\boldsymbol{n}}}}{\sqrt{1 + \tanh^2 \beta}}
    \label{eq:dissipative_steps_2}
\end{equation}
By changing back to the computational  basis
\begin{equation}
    \begin{aligned}
        \ket{\psi_2} = & \frac{1}{\sqrt{2}}\!\sum_{{s_a}=0,1} \!\! \left( \frac{1 +(-1)^{s_{a}} \Pn \tanh \beta}{\sqrt{1 + \tanh^2 \beta}} \right) \ket{s_a}_a\otimes \ket{\Omega_E}_{P_{\boldsymbol{n}}} \\
        \label{eq:dissipative_steps_3}
    \end{aligned}
\end{equation}
one can see how the measurement of the ancilla qubit in $\ket{0}_a$ projects the gauge qubits onto the desired variational state, which happens with $50\%$ probability. Conversely, if the state $\ket{1}_a$ is measured, the plaquette ends up in $(\cosh 2\beta)^{-1/2} \, {\rm exp}\{-\beta {P}_n\} \ket{\Omega_E}_{P_n}$. In this case, the correct variational state can be recovered by applying the operator ${X}_{(\boldsymbol{n}, i)}$  to one of the plaquette qubits. This operator flips the sign in front of the plaquette operator upon commutation, and consequently the sign in the exponential such that we deterministically obtain 
\begin{equation}
\ket{\psi_{\rm d}(\beta)}=   \frac{{\rm e}^{\beta {P}_{\boldsymbol{n}}}}{(\cosh 2\beta)^{1/2}} \ket{\Omega_E}_{P_{\boldsymbol{n}}}.
   \label{eq:dissipative_steps_4}
\end{equation}
We note that this can only be done because $\ket{\Omega_E}$ is an eigenstate of every ${X}_{(\boldsymbol{n}, i)}$ with eigenvalue $+1$, and is thus left invariant after the commutation, such that the non-unitary exponential can be implemented deterministically. We remark that no post-selection is necessary, which would imply discarding half of the experimental runs for a single plaquette. For $N_p$ plaquettes, the post-selection probability vanishes as $1/2^{N_{p}}$ in the thermodynamic limit $N_p\gg 1$, showcasing the difficulty of working with non-unitary ans\"{a}tze for large quantum many-body systems. In this respect, our deterministic protocol is crucial, as in the absence of errors it produces the target variational state with $100\%$ yield.

\begin{figure}
    \centering
    \includegraphics[width=0.99\linewidth]{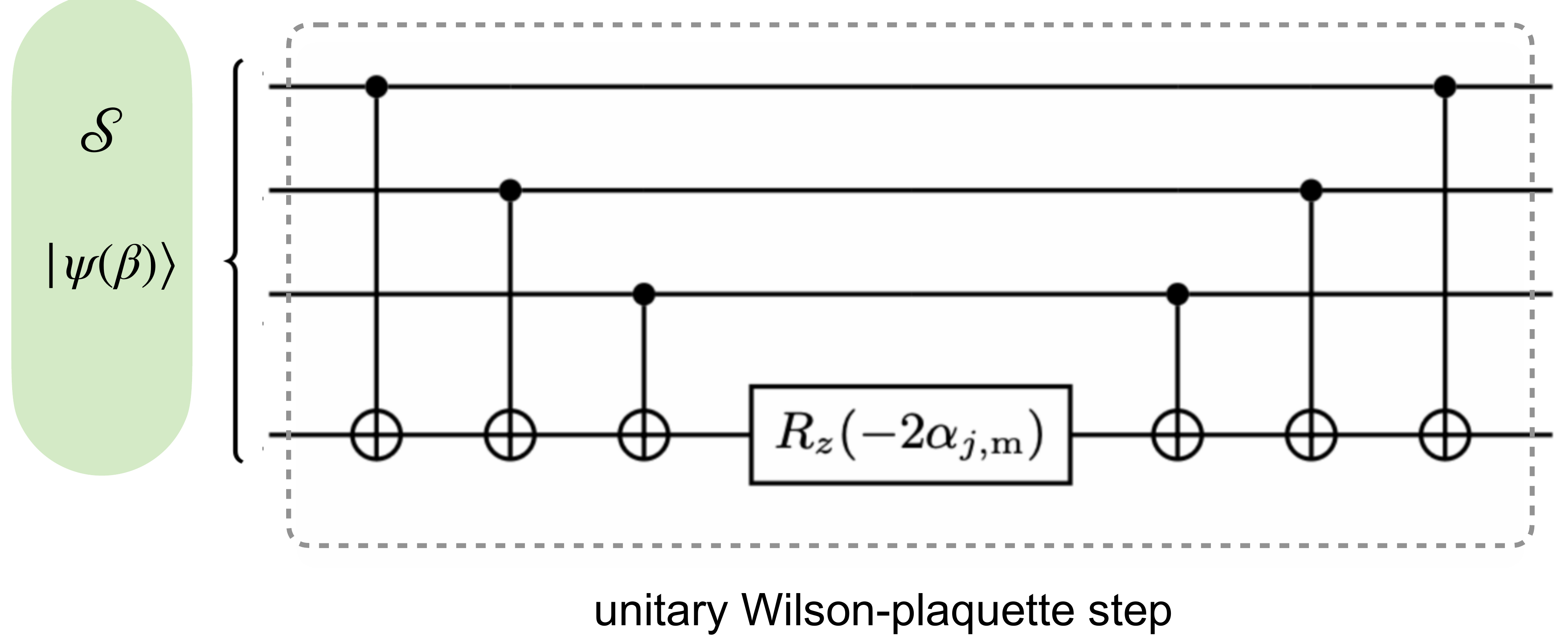}
    \caption{Circuit implementation of the unitary exponential of a single plaquette ${\rm e}^{{\rm i} \alpha_{j,{\rm m}} {P}_{\boldsymbol{n}}}$ considered in this work. This implementation does not use an extra ancilla qubit and thus requires an extended, but still local connectivity. An equivalent circuit making use of an ancilla and requiring only nearest-neighbor interactions can be found in Ref.~\cite{nielsen00}.}
    \label{fig:unitary_implementation}
\end{figure}

The complete non-unitary operation ${K}_{1, \beta}$, appearing in the variational ansatz~\eqref{eq:ansatz}, is realized by repeating the application of this circuit for each plaquette in the considered lattice. It is possible to do this in parallel using the following strategy: {\it (i)} Execute the unitary part (without the measurement) of the circuit of Fig.~\ref{fig:non_unitary_implementation} simultaneously for all plaquettes in the lattice. It is important to execute the train of CNOT gates in the correct order such that no physical qubit acts as the control of two CNOT gates simultaneously. This can be achieved, for example, by choosing the physical qubit in the top link of the plaquettes as the first control and then applying the remaining CNOT gates controlled on the other physical qubits in clockwise order. {\it (ii)} Measure all ancilla qubits in the computational basis, store the results in a classical register, and apply operators ${X}_{(\boldsymbol{n}, i)}$ to gauge qubits along strings which connect pairs of plaquettes whose ancilla qubits have been measured in state $\ket{1}_a$. An example is shown in Fig.~\ref{fig:plaquettes_correction}. These strings are required because applying the operator ${X}_{(\boldsymbol{n}, i)}$ to a qubit flips the sign of the exponential of all plaquettes in which that respective qubit is involved. The strings ensure that plaquettes whose sign does not need to be corrected are left invariant because they will receive either zero or two sign flips.
This procedure has the same overhead in terms of quantum computational resources as a single round of quantum error correction in the surface code~\cite{Terhal_2015}. The process of finding the set of strings that correct all plaquette signs is similar to the decoding process in the surface code. In this case, however, the existence of a solution is guaranteed because the prepared state is a $+1$ eigenstate of the operator acting as logical $X$-operator in the surface code.

The implementation of the unitary operations in Eq.~\eqref{eq:ansatz} has already been discussed in the literature. The exponential of the electric term of the Hamiltonian ${\rm e}^{{\rm i} \alpha_{j,{\rm e}} {H}_E}$ is a simple tensor product of single-qubit rotations along the $x$-axis, while the implementation of ${\rm e}^{{\rm i} \alpha_{j,{\rm m}} {P}_n}$ is a direct generalization of the circuits used to simulate the evolution of Ising interactions via CNOT gates and rotations about the $z$-axis ${\rm e}^{{\rm i}\alpha Z_cZ_t}={\rm CNOT} \, {\rm e}^{{\rm i}\alpha Z_t} \, {\rm CNOT}$, and can be found in the literature, e.g. see the circuits for the HVA in Refs.~\cite{burrelo, gustafson2023quantum}. Figure~\ref{fig:unitary_implementation} shows a possible implementation for the unitary exponential for a single plaquette, ${\rm e}^{{\rm i} \alpha_{j,{\rm m}} {P}_{\boldsymbol{n}}}$. The scheme requires the sequential application of CNOT gates between qubits of a plaquette. The gates can be chained such that only interactions between neighboring gauge qubits are required, which is reasonable for e.g. neutral atom platforms~\cite{morgado_2021_quantum}. For devices in which interactions only reach from an ancilla qubit to the qubits of the respective plaquette, (i.e.~nearest-neighbor connectivity), the ancilla can be utilized to implement the required operations~\cite{nielsen00}. The application of the unitary plaquette exponentials can also be parallelized~\cite{burrelo}.

A complete variational circuit for the $d=3$ lattice and two variational layers is shown in Fig.~\ref{fig:full_circuit} in Appendix~\ref{app:circuit}. Parallelizing all operations yields a circuit depth of $13$ per unitary layer. Together with a circuit depth of $9$ for the first variational layer of the non-unitary operator and the electric-field unitary, yields a total circuit depth of
\begin{equation}
    D = 13\ell_u + 9,
    \label{eq:circuit_depth}
\end{equation}
where $\ell_u$ is the number of unitary layers. Note that the circuit depth is independent of the lattice size. Single and two-qubit gates as well as measurements count as one unit of depth. The parallel execution of gates on independent qubits also contributes to one unit of depth. Another relevant quantity to evaluate the complexity of the dissipative variational algorithm is the number of CNOT gates required to prepare the variational state, since these are the main source of circuit-level noise in real platforms. For the boundary conditions considered in this paper, the number of CNOT gates amounts to
\begin{equation}
    \# \mathrm{CNOTs} = (d_x - 1) d_y (4 + 6 \ell_u) - (d_x - 1) (2 + 4 \ell_u) .
\end{equation}
The fact that the dissipative variational ansatz can achieve good performance with short circuit depths $\ell_u$ is crucial for maintaining the feasibility of the variational algorithm on real hardware, where the algorithms are essentially limited by this metric being low enough. In the next section, we show that high precision in the ground-state energy can be achieved already with a single variational layer. The depth of the corresponding circuit is certainly within the range of what current NISQ devices can achieve, as the short depth avoids the proliferation of errors. The analysis of the ground-state preparation algorithm in the presence of noise in Sec.~\ref{sec:results_noise} reveals that a small gate error rate is required for the introduction of the first unitary layer not to be counterproductive.

\section{\bf Noiseless dissipative VQE: A non-unitary advantage} \label{sec:noiseless_results}

\begin{figure}[t]
    \centering
    \hspace*{-0.8cm}
    \includegraphics[width=1.05\linewidth]{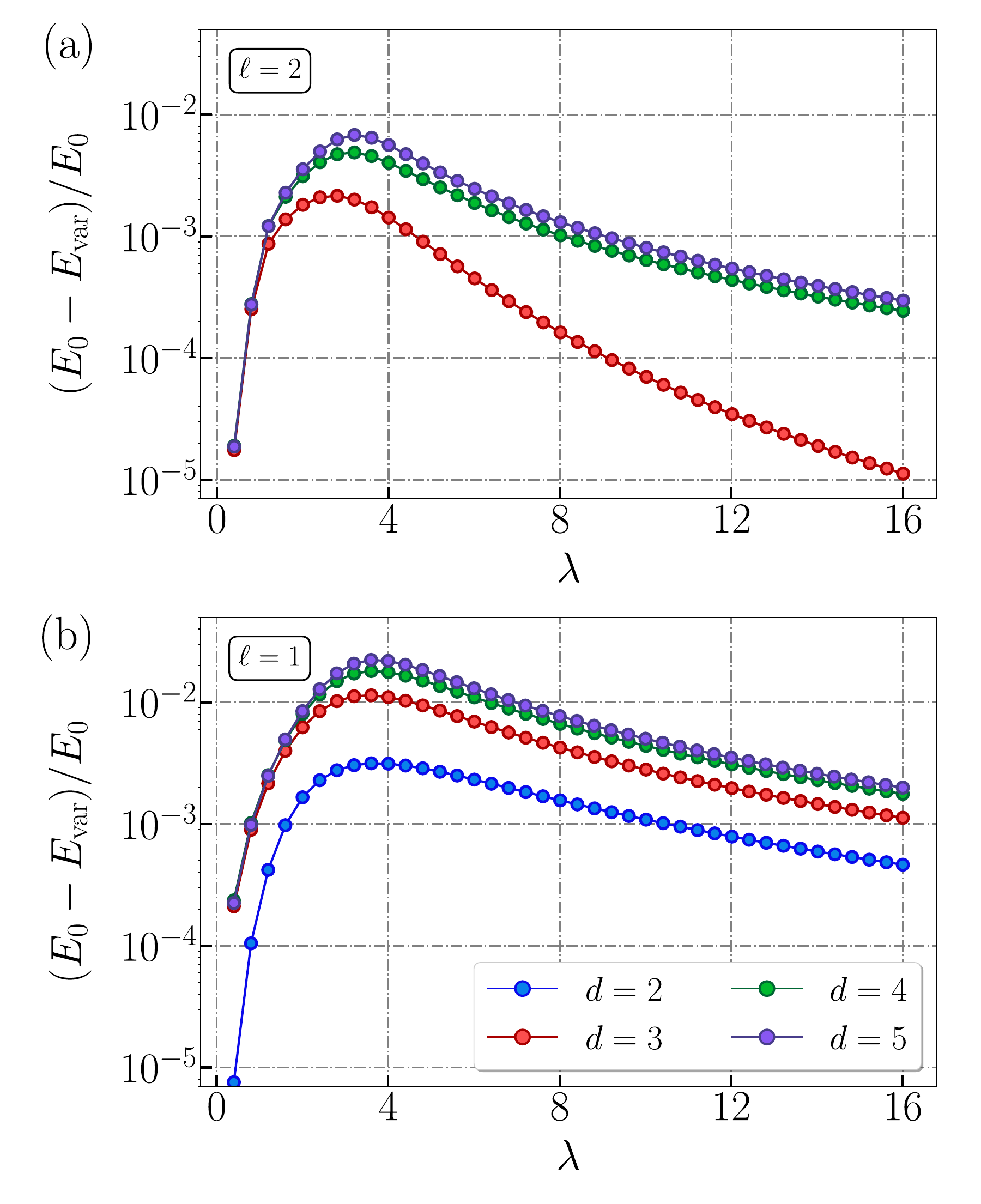}
    \caption{Ansatz error quantified by the relative energy difference between the minimum variational energy and the true ground-state energy obtained from exact diagonalization. We display this relative error for different values of $\lambda$ as the lattice size increases with the distance $d$, considering two different numbers of variational layers $\ell\in\{1,2\}$. For $d=2$, the $\ell=2$ ansatz is expressive enough to achieve the exact energy up to numerical precision.}
    \label{fig:noiseless_ansatz_performance}
 \end{figure}

We now present results from classical state-vector simulations of the ground-state preparation algorithm to analyze the performance of the ansatz. We have performed simulations investigating the ansatz in the absence and presence of circuit-level noise. In this section, we focus on the former case. We have restricted the Hilbert space of noiseless simulations to that defined by constraint \eqref{eq:gauge_invariance}, to reduce the dimension of the Hilbert space from $2^{N}$ to $2^{N_p}$, so that we gain access to larger lattices that are beyond a brute-force exact-diagonalization approach (up to $d=5$, 41 qubits). We work in the basis \eqref{eq:plaquette_basis}. We first present our results for the energy difference between the DVA  and the exactly diagonalized ground state, which can be interpreted as a physical figure of merit quantifying the error of the various variational approximations (Fig.~\ref{fig:noiseless_ansatz_performance}). We then present data for the value of the dual magnetization, which is the relevant non-local order parameter that can capture the nature of the confinement-deconfinement transition. As discussed in more detail in  Appendix~\ref{app:Z2}, the dual magnetization is defined along a path $\partial \mathcal{C}_n$  that extends from its boundary to one of the links  in plaquette ${P}_{\boldsymbol{n}}$, as depicted in   Fig.~\ref{fig:title}(a), and reads
\begin{equation}
    {M}_n = \prod_{(\boldsymbol{n}, i) \in \partial \mathcal{C}_n} {X}_{(\boldsymbol{n}, i)}.
    \label{eq:dual_magnetization}
\end{equation}
Even with a short circuit depth $\ell = 2$, the DVA \eqref{eq:ansatz} can accurately account for this dual magnetization in the largest lattice, providing a reasonable estimate of the critical exponents of the $\mathbb{Z}_2$ LGT, which quantifies its performance in the critical region of the model. We repeat the same analysis with the unitary ans\"atze \eqref{eq:HVA} and \eqref{eq:HVA_B}, and show that they exhibit large discrepancies.

\begin{figure*}
    \centering
    \includegraphics[width=0.85\textwidth]{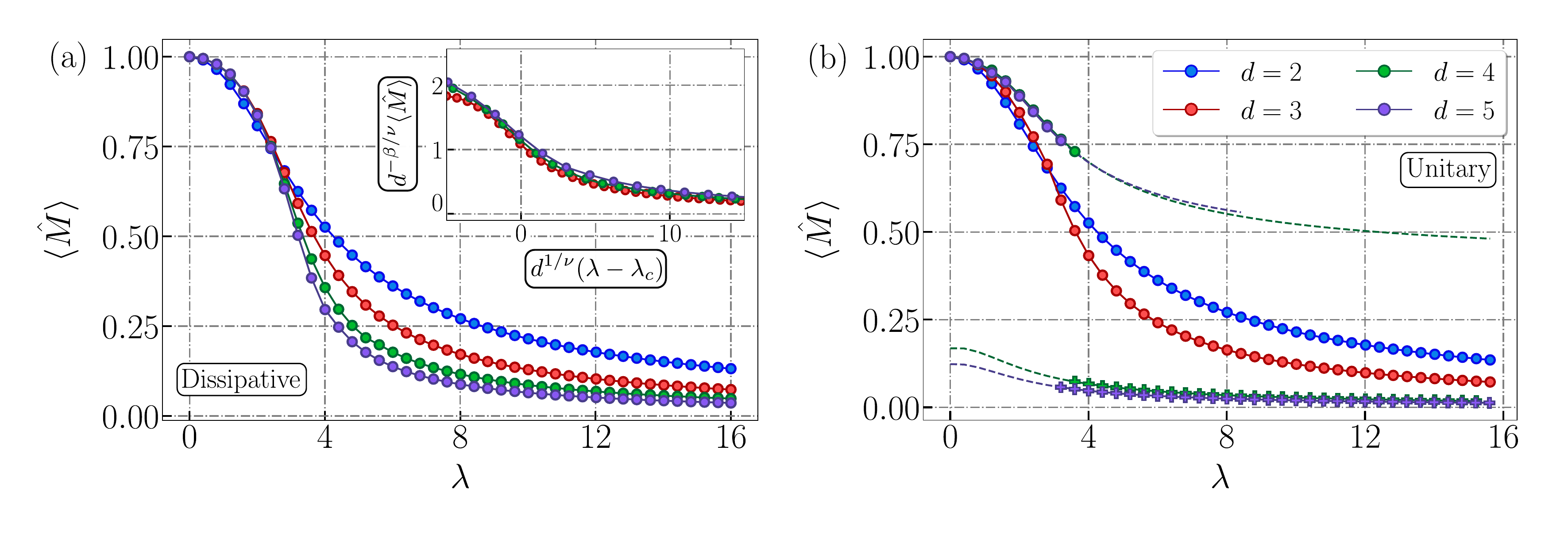}
    \caption{Dual magnetization averaged over all plaquettes in the bulk of the lattice ${M} = N_{p, \mathrm{bulk}}^{-1} \sum_{n \in \mathrm{bulk}} {M}_n$ as a function of $\lambda$ predicted by the $\ell = 2$ dissipative (a) and unitary (b) variational ans\"atze with the same depth: ($\ket{\phi_\mathrm{u, e}}, \ell_u = 2$) and ($\ket{\phi_\mathrm{u, m}}$, $\ell_u = 1$). We are considering the minimum depth required to prepare the reference state $\ket{\Omega_B}$ for the $\ket{\phi_\mathrm{u, m}}$ ansatz which would require dissipation. This extra overhead results in one less unitary variational layer to maintain equal depth. The inset in (a) shows the curve collapse after the finite-size scaling analysis (see Appendix~\ref{app:curve_collapse} for details). In (b), circles (crosses) represent data obtained using the electric (magnetic) unitary ansatz  $\ket{\psi_{\rm u,e}}$ ($\ket{\psi_{\rm u,m}}$). Dashed lines represent the continuation of the respective dual magnetization curves into the phase where the respective unitary ansatz does not yield the minimum energy among the two. Large discrepancies are observed between the two ans\"atze, showcasing the superiority of the proposed dissipative approach for few variational layers.}
    \label{fig:magnetizations}
\end{figure*}

All the following results rely on our classical optimization strategy to estimate the optimal variational parameters. Even though we use a standard gradient descent algorithm \cite{gradient_descent}, we introduce an initialization strategy that is in some sense inspired by adiabatic evolutions. Appendix~\ref{app:optimization_algorithm} describes such an optimization strategy. In Appendix~\ref{app:parameter-shift-rule}, we formulate an extension of the parameter-shift rule \cite{parameter_shift_og, parameter_shift_good} for the exact computation of gradients. We find that our optimization strategy is stable, yielding the same results for different runs, and also showing a better performance than random initialization.

Figure~\ref{fig:noiseless_ansatz_performance} shows the relative difference between the energy obtained by the DVA  \eqref{eq:ansatz} and the exactly diagonalized ground-state for increasing lattice sizes. We have considered only small total depths $\ell = \ell_u + 1 \in\{ 1, 2\}$ since, as discussed in Sec.~\ref{sec:results_noise} in more detail when considering imperfect gates in an experiment, extremely small error rates would be required for the introduction of more layers to be beneficial. In fact, the strength of the dissipative ansatz is that it can converge to the true ground state energy with few layers. The $\ell = 2$ ansatz can consistently achieve relative energy differences well below the percent level for every value of $\lambda$, and for lattice sizes as large as $d = 5$ ($N=41$ qubits). We remark that the use of a dissipative layer allows us to interpolate between the two different phases of the gauge theory using only a single reference state. In Appendix~\ref{app:diss_vs_unitary}, we show that the relative energy has a similar profile to the state infidelity, which is a typical figure of merit used in the variational method. There, we also compare the relative energies achieved by the unitary approach of Ref.~\cite{burrelo} and our ansatz. This, combined with the results of Sec.~\ref{sec:results_noise} shows that the dissipative variational ansatz achieves considerably lower energies for the shallow depths accessible with current and near-term devices, especially around the critical point of the model. In this respect, the performance of the shallow dissipative variational ansatz requires focusing on physical quantities that encode the relevant features of the confinement-deconfinement transition. It is well known that state fidelity has a limited predictive value in the thermodynamic limit of a many-body system \cite{PhysRevLett.18.1049}. Likely, we note that the energy of a certain variational ansatz can show a promising convergence to the true value even when the state poorly reflects the critical properties of the model.

\begin{table}[t]
    \centering
    \begin{tabularx}{0.47\textwidth}{cccccc}
        & \,\,\,\,\,Dissipative  & \,\,\,\, Monte  & \,\,\,\,\, Exact  & \,\,\, Unitary & \,\,\, Unitary \\
        & \,\,$\ket{\psi_{\rm d}}$~\eqref{eq:ansatz} &  \,\,\,Carlo &  \,\,\,\,diag. & $\,\,\ket{\phi_{\rm u,e}}$~\eqref{eq:HVA} &  \,\,$\ket{\phi_{{\rm u,m}}}$~\eqref{eq:HVA_B}\\ \hline \hline
        $\lambda_c$ & 3.24 & 3.04 & 3.06 & 2.56 & 2.09 \\
        $\beta$ & 0.35 & 0.33 & 0.36 & 0.04 & -1.27 \\
        $\nu$ & 0.59 & 0.63 & 0.64 & -0.20 & -0.40
    \end{tabularx}
    \caption{Comparison between the critical exponents predicted by the DVA and HVA ans\"atze with $\ell=2$ variational layers (see Fig.~\ref{fig:magnetizations}), exact diagonalization of the systems up to lattice size $d=5$, and Monte Carlo simulations~\cite{critical_lambda, ising_critical_exponents}.}
    \label{tab:critical_exponents}
\end{table}

Therefore, we now turn our attention to the estimation of the dual magnetization~\eqref{eq:dual_magnetization}. In Fig.~\ref{fig:magnetizations}(a), we show data obtained from the DVA with $\ell = 2$ variational layers. This curve shows the expected behavior for the dual magnetization, which takes values close to 1 in the confined phase, decreasing continuously around the phase transition, and eventually approaching 0 for increasing $\lambda$ in the deconfined phase. The magnetization does not completely vanish in the latter regime because we are considering reduced lattice sizes, and boundary effects become manifest. However, one can see how, as one increases the system size, the decay becomes more pronounced such that, in the thermodynamic limit, one can obtain the characteristic non-analyticity of a second-order quantum phase transition. Comparison with Fig.~\ref{fig:magnetizations}(b) shows that the unitary ansatz with the same depth is not able to reproduce this behavior, not even if the long-range entangled state $\ket{\Omega_B}$~\eqref{eq:magnetic_gs} is used as a reference state~\eqref{eq:HVA_B}. We remark that, by choosing the respective reference state with the lowest-energy HVA for a specific value of $\lambda$, one obtains a discontinuous magnetization curve as the lattice size increases, which is contrary to the continuity expected from a second-order phase transition. 

On the other hand, if one sticks to a single reference state all along the critical point following the corresponding dashed lines of Fig.~\ref{fig:magnetizations}(b), continuity is maintained but large discrepancies appear in the predicted criticality, as we now outline. To quantify these, we perform a finite-size scaling analysis to estimate the critical exponents of the $\mathbb{Z}_2$ LGT with different methods. Table~\ref{tab:critical_exponents} gives a summary of the various results, offering a comparison between the dissipative and unitary ans\"atze, exact diagonalization for the same lattice sizes and boundary types, as well as state-of-the-art Monte Carlo results that serve as a guide to quantify the performance of each of these approaches.
In Appendix~\ref{app:curve_collapse} we outline the details of the finite-size scaling analysis used to extract these results, which clearly show the large deviations between the predictions of the dissipative and unitary variational ans\"atze. The scaling analysis of the DVA in the critical region is stable for constant depths, giving values of the critical exponents that are close to the expected ones. The reason is that including the dissipative dynamics into the ansatz allows it to adaptively generate different amounts of entanglement. The unitary ansatz is less effective in this respect, and its regime of accuracy lies well inside each of the two phases of the model, where the entanglement present in the true ground state is more similar to that of the reference state. In the critical regime, a sudden change in the entanglement present in the ground state causes the shallow unitary ansatz to lose its predictive power about the confinement-deconfinement phenomenon, while the dissipative one is capable of capturing the essence of this transition with constant-depth quantum circuits. In NISQ devices, due to the accumulation and propagation of gate errors,  the maximum number of layers for a given qubit number will face practical limitations. As discussed above, a low-depth  unitary ansatz struggles to capture the sudden changes characteristic of phase transitions. This is generic for HVAs which, according to generic Lieb-Robinson bounds, would require a layer number that grows with  the system size to generate the required entanglement. Hence, sticking to a constant-depth HVA will limit the achieved accuracies. In our case, it prevents the dual magnetic susceptibility from diverging as one would expect from a second-order phase transition. This divergence fixes the value of $\nu$ (see Appendix \ref{app:curve_collapse}), which is involved in the computation of every other critical exponent. We observe that for $\ell = 2$, this exponent becomes negative, and in consequence, the results from the finite size scaling analysis using the unitary ansätze are far from the more precise Quantum Monte Carlo data.

In Appendix~\ref{app:diss_vs_unitary} we also compare the energy achieved with the DVA ~\eqref{eq:ansatz} and the completely unitary variational ans\"atze~\eqref{eq:HVA} for increasing number of layers in the $d=4$ lattice. There we show that a completely unitary ansatz starting from a product state is not able to cross the phase transition, even with depths up to $\ell = 4$. In Appendix~\ref{app:Z2}, we show the DVA data for the Creutz ratio and the topological entanglement entropy, two quantities that behave as expected.

\section{\bf Noisy dissipative VQE: a layer-number threshold} \label{sec:results_noise}

NISQ devices are intrinsically noisy and no platform has yet been scaled up to the point of achieving practical fault-tolerant quantum computation. This limits the maximum depth of feasible quantum algorithms, as longer algorithms lead to the accumulation of more physical errors. Variational quantum circuits consisting of more layers are in principle more expressive and yield a better approximation of the true ground state for the $\mathbb{Z}_2$ LGT and physical quantities calculated from that state. On NISQ hardware, however, one has to find a balance between a circuit that is shallow enough to not induce too much noise and, at the same time, sufficiently deep to prepare an accurate variational state. In this section, we investigate the effect of noise in our dissipative VQE proposal for the $\mathbb{Z}_2$ LGT. To quantify it, we use the increase in the variational ground-state energy w.r.t the value that is achieved in the noiseless simulation as a metric. We also discuss a strategy for detecting the presence of errors when measuring the energy of the variational state that allows the detection of some particular Pauli errors that can occur during the ground-state preparation process.

As noted in Sec.~\ref{sec:circuit_implementation}, the pure $\mathbb{Z}_2$ LGT is closely related to the surface code for quantum error correction. Given a lattice of qubits as shown in Fig.~\ref{fig:title}(a), the subspace spanned by the $+1$-eigenstates of all gauge operators $\Gn$ and plaquette operators $\Pn$ has dimension two and can therefore encode a single logical qubit. In the context of quantum error correction, the gauge and plaquette operators are referred to as $X$- and $Z$-stabilizers, respectively. $X$-stabilizers are measured to detect $Z$-errors on a code state and vice versa. Any measurements that yield the value $-1$ indicate the presence of errors. The set of all stabilizer measurement outcomes, known as the error syndrome, is then used to determine a correction to be applied on the code state \cite{Terhal_2015}.

\begin{figure*}[t]
    \includegraphics[width=0.85\linewidth]{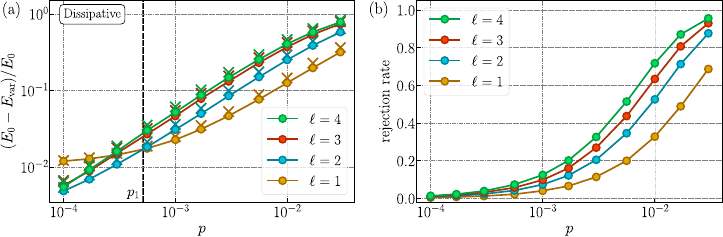}
    \caption{Variational ground-state preparation in the presence of noise: We consider the $d=3$ lattice (13 data qubits and 6 ancilla qubits) at $\lambda = 3.0$. Variational states are prepared by applying $\ell$ layers of the proposed DVA in the presence of depolarizing circuit-level noise of strength $p$. (a) Precision of the estimated ground-state energy: Circles correspond to post-selected data while crosses mark the raw data. With a decreasing error rate, it becomes beneficial to apply more variational layers --- the first layer-number threshold $p_1 \approx 5 \times 10^{-4}$ is marked by a vertical line. State-of-the-art two-qubit gate error rates lie roughly at $p_{2q} \approx 3 \times 10^{-3}$, which means that the application of a single variational layer yields the most precise ground-state energy estimate. (b) Rejection rates in the post-selection process. When all qubits are measured in the $X$-basis, the measurement values can be used to calculate the values of the gauge operators $\Gn$. Whenever one of these observables is measured to be $-1$ we reject the run. As expected, the rejection rates are higher for more variational layers since the probability for errors to occur grows with the circuit depth.}
    \label{fig:noisy_GS_energy_diss_rejection}
\end{figure*}

States that are $+1$-eigenstates of all $X$-stabilizers of the surface code are valid physical states in the pure $\mathbb{Z}_2$ LGT. Since the variational ansatz investigated in this work fulfills the gauge constraints defining the physical subspace~\eqref{eq:physical_states}, the eigenvalue of all gauge operators can be measured to detect and correct $Z$-errors arising during the ground-state preparation process. Moreover, this can be done in parallel to the measurement of the energy of the variational ground-state, since the terms contained in the electric term of the Hamiltonian are also involved in the expectation value of gauge operators. However, the correction of such errors is more subtle than in the surface code, where $Z$-errors need only be corrected up to a plaquette operator $\Pn$. Unlike in the surface code, physical states of the pure $\mathbb{Z}_2$ LGT are not restricted to be $+1$-eigenstates of all plaquette operators. In general, they are not even eigenstates of the plaquettes which means that the application of an operator $\Pn$ produces a different valid physical state with higher energy than the ground state. Thus, $Z$-errors on variational states can only be corrected if no other error of the same weight yields the same pattern of violated gauge operators, i.e. the same error syndrome. In contrast to the surface code, $X$-errors cannot be detected in the $\mathbb{Z}_2$ LGT by measuring plaquette operators, since plaquettes are only stabilizers in the limiting case $\lambda \to \infty$. $X$-errors map physical states onto other physical states and are thus impossible to correct within our dissipative VQE. We note that for the boundary conditions considered in this work (Fig.~\ref{fig:title}(a)), the gauge operators $\Gn$ involve at least 3 qubits which means that any weight-2 $Z$ error is detectable in lattices with size $d \geq 3$. Moreover, every weight-1 $Z$ error has a unique syndrome and is therefore correctable, as discussed above. In the so-called rotated surface code, certain stabilizers involve only 2 qubits, and not every weight-1 error causes a unique syndrome \cite{bombin2007optimal}. Therefore, if the $\mathbb{Z}_2$ LGT is implemented on the lattice of the rotated surface code, there exist uncorrectable weight-1 $Z$ errors and weight-2 $Z$ errors that are not detectable.

In the following, we analyze the effect of circuit-level noise during the preparation of variational states on the estimated $\mathbb{Z}_2$ LGT ground-state energy. We perform numerical state vector simulations using the Python package PECOS \cite{ryan2018quantum, pecos}. We note that the energy minimization was performed in the absence of noise to avoid introducing extra overhead in the already highly non-convex optimization problem \cite{pellowjarman2023qaoa}. The noisy simulations are then performed utilizing the optimal variational parameters found after that process. In the simulations, we employ depolarizing noise, which is the most general noise model for computational errors.
The depolarizing channel of strength $p$ reads 
\begin{equation}
	\mathcal{E}_p(\rho) = (1-p)\rho + \frac{p}{4^q-1} \sum_{i} {\mathcal{P}}_i \rho {\mathcal{P}}_i , \label{eq:depol}
\end{equation}
with ${\mathcal{P}}_i \in \{{I},\,{X},\,{Y},\,{Z}\}^{\otimes q} \backslash \{{I}^{\otimes q}\}$ and $q$ the number of qubits the noise channel acts upon. In systems with long coherence and relaxation times, errors mainly occur during the application of gates. In our simulations, every gate is thus followed by possible faults whereas we do not consider errors on idling positions. Specifically,
\begin{itemize}
    \item A single-qubit gate is followed by a Pauli fault drawn uniformly and independently from $\{{X},{Y},{Z}\}$ with probability $p/3$.
    \item A two-qubit gate is followed by a two-Pauli fault drawn uniformly and independently from $\{{I},{X},{Y},{Z}\}^{\otimes 2} \backslash \{{I} \otimes {I}\}$ with probability $p/15$.
    \item Qubit initialization is flipped ($\ket{0} \mapsto \ket{1}$) with probability $2p/3$.
    \item Qubit measurements yield a flipped result ($\pm 1 \mapsto \mp 1$) with probability $2p/3$.
\end{itemize}

We refer the reader to Appendix~\ref{app:optimization_algorithm}, where more details regarding the noisy simulations can be found. In particular, we start with the state $\ket{0}^{\otimes N}$ and first apply Hadamard gates to all qubits encoding the gauge degrees of freedom to obtain the ground-state $\ket{\Omega_E}$ of the electric Hamiltonian (cf. Eq.~\eqref{eq:elec_term}). We then apply variational layers, as shown in Fig.~\ref{fig:full_circuit} of Appendix~\ref{app:circuit} for the $d=3$ lattice. After the state preparation, we measure all qubits individually in either the $X$-basis or the $Z$-basis. When measuring in the $X$-basis, it is possible to deduce the values of the gauge operators and post-select onto states that fulfill all gauge constraints. By doing so, we can sort out states in which detectable $ Z$ errors occurred.  Repeating the preparation procedure followed by measurements in those two bases allows us to calculate the energy of the variational state according to Eq.~\eqref{eq:hamiltonian}. In Fig.~\ref{fig:noisy_GS_energy_diss_rejection}(a), we compare the estimated energies of several variational states on the $d=3$ lattice in the presence of noise. We choose $\lambda = 3.0$ such that the system is in the confined phase but very close to the phase transition at which it is most difficult to get good estimates for the ground-state energy. One can see an expected linear scaling for large error rates. For small values of $p$, the curves approach values corresponding to the energy precision which can be reached with the respective number of layers in the absence of circuit noise.
\begin{figure*}[t]
    \includegraphics[width=0.85\linewidth]{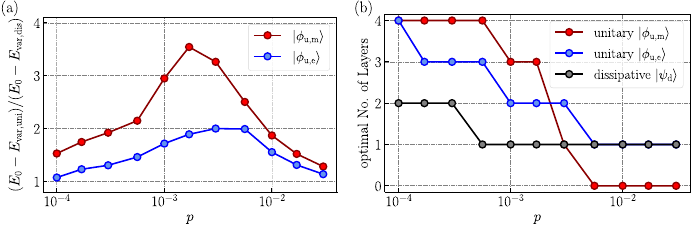}
    \caption{Comparison of dissipative and unitary variational state preparation in the presence of circuit-level noise of strength~$p$. Ground-state energies are estimated for the $d=3$ lattice (13 data qubits and 6 ancilla qubits) at $\lambda = 3.0$. (a)~Advantage of dissipative state preparation w.r.t.~both unitary state preparation schemes. The number of variational layers $\ell$ is chosen such that the lowest possible energy is obtained, see panel (b). Since we examine the LGT close to the phase transition, we investigate the unitary state preparation starting from the trivial reference state, resulting in $\ket{\phi_{\rm u, e}}$, as well as the surface code reference state, which gives $\ket{\phi_{\rm u, m}}$. The preparation of the surface code state $\ket{\Omega_B}$ is conducted by measuring all plaquette operators in the presence of circuit-level noise. At state-of-the-art two-qubit gate error rates $p_{2q} \approx 3 \times 10^{-3}$ the advantage of dissipative variational state preparation is the largest. (b) Optimal number of variational layers as a function of the error rate $p$. For current gate error rates the DVA requires fewer variational layers, resulting in a lower sensitivity to noise.}
    \label{fig:noisy_comparison_diss_uni}
\end{figure*}
It becomes evident from the plot that, for low error probabilities, it is beneficial to apply more variational layers. On the other hand, for realistic noise strengths,  circuits consisting of fewer layers are advantageous as there is a lower accumulation of errors. In addition, as shown in Fig.~\ref{fig:noisy_GS_energy_diss_rejection}(b) where we display the rejection rates associated with the aforementioned post-selection process using up to 4 variational layers, errors are more likely to occur in deeper circuits with more layers, resulting also in larger rejection rates and raising the practical complexity of an experiment. We can now give a specific noise-aware assessment of the optimal variational strategy in the NISQ regime, going in this way beyond the typical theoretical discussion that focuses on the expressivity of a variational ansatz and limitations due to Barren plateaus in the limit of a large number of layers. We find that well before this phenomenon becomes a limitation, one faces the fact that increasing the number of layers leads to worse variational estimates. We find that,  for error rates $p$ larger than $p_{1} \approx 5 \times 10^{-4}$, the single-layer variational ansatz yields better energy estimates than those with a larger number of layers due to its small circuit depth and, consequently, its reduced sensitivity to errors. For our simple noise model,  error rates smaller than  $10^{-4}$ are required for an ansatz with more than two variational layers to become advantageous. We note that in state-of-the-art neutral atom quantum processors, single-qubit gate infidelities of $p_{1q}\approx3\times 10^{-4}$ have been demonstrated and two-qubit gates were realized with error rates of $p_{2q} \approx 5 \times 10^{-3}$~\cite{evered2023highfidelity}. In trapped ion platforms, single- and two-qubit gate infidelities of $p_{1q} \approx 7 \times 10^{-5}$ and $p_{2q} \approx 3 \times 10^{-3}$, respectively, have been reported in \cite{ryananderson2021realization}. Since the majority of gates in the variational circuit are CNOT gates, we expect the performance to be limited by the two-qubit gate fidelity. For a small layer number, the limitations of the DVA could also be dominated by measurement errors, if these were noisier than the CNOT gates. In any of these cases,  the single-layer variational ansatz is expected to give the most precise ground-state energy estimates in currently available hardware platforms. As hardware and software advances improve the gate fidelities, one will enter regimes for which the application of more than one variational layer becomes beneficial within NISQ. Despite the present limitations, we would like to highlight that our proposed DVA does not require a large number of layers to achieve considerable accuracy in the prediction of the critical nature of the confinement-deconfinement phase transition. As shown in the previous section, $\ell=2$ layers suffice to give competitive predictions for systems with  $N=41$ qubits. Given the increased complexity of classical simulations of 2d models, it is interesting to note that some NISQ devices already operate with hundreds of physical qubits, and could soon go beyond these numbers reaching for instance $d=7$ and $d=9$. It should also be noted that, in contrast to other near-term applications and use cases that require considerably large numbers of $T$ gates~\cite{scholten2024assessing}, our circuits only require a reduced amount of non-Clifford operations corresponding to the single-qubit rotations for the ancilla and physical qubits (roughly scaling with the qubit number, e.g. $N_{\rm nc}=38$ gates for  $d=3$ and $N_{\rm nc}=74$ for $d=5$) which, depending on the target error rate, would have additional overhead in terms of $T$ gates~\cite{10.5555/2685188.2685198}. Therefore, they also provide an interesting application with a reasonable resource complexity for the first application of small circuits of quantum-error-detected or corrected logical qubits, which should allow one to decrease the logical error rate concerning these values of $p$ by increasing the redundancy. It is in this context of logical qubits and mid-term small logical circuits that the non-Clifford gate counts are a relevant resource estimation.

In Fig.~\ref{fig:noisy_comparison_diss_uni}, we compare the ground-state energy estimates for the $d=3$ lattice at $\lambda = 3.0$ obtained with the DVA to the estimates from the Hamiltonian variational ansatz in the presence of circuit-level noise. For the HVA we investigate both cases, the one that starts the variational state~\eqref{eq:HVA} preparation from the electric-field dominated state $\ket{\Omega_E}$ and the one which first prepares the surface-code magnetically-dominated state $\ket{\Omega_B}$ and then applies the corresponding unitary layer~\eqref{eq:HVA_B}. Panel (a) exhibits the ratio between the energy precision obtained with each unitary ansatz and the dissipative one. We observe that, for state-of-the-art gate error rates, the DVA always yields more precise energies than any of the unitary ans\"atze,  as evidenced by the fact that the curves always lie above 1. This advantage in the presence of noise becomes more pronounced around the error rates reported in various NISQ devices, which thereby implicitly shows that the DVA has a higher expressivity for a few variational layers even in the presence of noise. We note that for $p \rightarrow 0$, arbitrary numbers of variational layers can be applied and thus, non-surprisingly, the advantage of the dissipative scheme to unitary ones regarding the energy reduces. On the other hand, in this very regime where errors are absent, we have shown in the previous section that one must look beyond the ground-state energy to assess the adequacy of the ans\"{a}tze to capture the confinement-deconfinement transition. From that perspective, the dissipative approach has a clear superior performance (see Table~\ref{tab:critical_exponents}). More detailed data on the energy precision of the variational states prepared with the HVA can be found in Fig.~\ref{fig:noisy_GS_energy_uni} of Appendix~\ref{app:diss_vs_unitary}, which gives further evidence that good ansatz expressivities at short circuit depth are crucial for NISQ devices, such that the dissipative approach should be preferred to the unitary one. We note that the simulation of noisy state preparation is computationally expensive, and thus we are limited to considering the $d=3$ lattice for our investigation, which already involves state vector simulations of 19 qubits. In Sec.~\ref{sec:noiseless_results} we showed that the advantage of the dissipative VQE for a fixed small number of variational layers grows with the system size. Thus, we expect that also in the presence of noise, the advantage of the proposed ansatz gets more pronounced for larger lattice sizes.

Let us now turn to a more detailed account of the optimal layer numbers for VQEs in the presence of noise. We have seen that there is a specific physical error rate below which increasing the layer number from $\ell=1$ to $\ell=2$ becomes advantageous. This is a particular instance of a variational error threshold, which determines the error rates $p_{L}$, below which it would be beneficial to increase the number of layers $\ell$. In panel (b) of Fig.~\ref{fig:noisy_comparison_diss_uni} and in Appendix~\ref{app:diss_vs_unitary}, we can see how these variational error thresholds appear for all the different unitary and dissipative ans\"{a}tze studied in this work, which corresponds to the particular error rates in which each of the curves shows a plateaux change. We believe that the concept of variational thresholds should become an important feature in the study of VQEs in the presence of noise.  For instance, considering error rates slightly below the percent level, Fig.~\ref{fig:noisy_comparison_diss_uni}(b) shows that the unitary HVA ansatz that starts from the magnetic-field dominated state should not consider any unitary layer. The best it can do is simply prepare the surface-code reference state, which will be incapable of capturing any criticality of the confinement-deconfinement transition. For this range of error rates, only the dissipative and unitary electric ans\"{a}tze can have a layer, the dissipative one is superior in terms of the attainable relative error.

In summary, in light of all these results,  the most accurate ground-state energy estimates can be obtained by applying just a single dissipative variational layer for state-of-the-art gate error rates. Future hardware improvements can make the application of two variational layers beneficial. Therefore, on current and near-future hardware devices the dissipative VQE yields the most accurate results.

\section{\bf Conclusions and outlook}

In this article, we proposed a combined dissipative-unitary approach for the study of ground-state properties of the $\mathbb{Z}_2$ LGT in QSs. We have shown that this algorithm can outperform previous unitary variational eigensolvers, especially when scaling to larger system sizes and around the phase transition of the LGT and already apparent with moderate lattice sizes. Note that we use a single variational ansatz that interpolates between the confined and deconfined phases to study both phases of the theory and its critical point, contrary to the unitary HVA. The resources required for the implementation of our scheme are in close correspondence with the capabilities of current NISQ devices, and the geometry of the two-dimensional $\mathbb{Z}_2$ LGT  leads to very natural mappings onto devices in which qubits are arranged in two-dimensional lattices, providing both a nice benchmark for the simulators themselves and an opportunity to explore exotic physics with these tools. We provided a detailed analysis of the variational ansatz prediction for quantities of physical interest, from which the study of confined-deconfined phase transitions and topological phases in QSs can benefit. Furthermore, we conducted a finite-size scaling analysis, showing that the introduction of dissipative dynamics is beneficial, if not required, in VQE studies of critical phenomena in these devices. Our analysis of the ground state preparation process in the presence of noise shows that the good ground state energy estimates, at small circuit depths achievable with our ansatz, persist in noisy hardware at current gate error rates, extending the above advantage to realistic accounts of NISQ devices. We also discuss a post-selection scheme that can be used to discard faulty states when measuring the energy and other observables which are tensor products of ${X}_{(\boldsymbol{n}, i)}$-operators.

The use of techniques related to imaginary time evolution in QSs is currently a very active area of research. In future work, it would be interesting to explore whether one can extend the proposed combined dissipative-unitary approach to more complex lattice gauge theories, either containing matter degrees of freedom or more complex gauge groups. Another open question is the preparation of excited states which could serve as a starting point for simulations of dynamic scenarios. Concerning the application of our proposal on real hardware, it would be interesting to study the parameter optimization in the presence of circuit-level errors as well as shot noise in the measurements. Future work could also include the investigation of more sophisticated noise models, such as biased noise. Moreover, it is an open question whether techniques from fault-tolerant quantum computing can be applied to directly reduce the effect of errors during the preparation of variational states.

\section*{Code availability}
All codes used for data analysis are available from the authors upon reasonable request.

\section*{Author contributions}
A.B. and E.R. conceived the initial idea about the use of non-unitary VQEs for LGTs and, together with M.M., considered its ancilla-assisted circuit implementation. J.C. and E.R. devised the optimal variational scheme combining both unitary and non-unitary layers, as well as its deterministic implementation. D.L. and M.M. designed the ideas and methods to understand the effect of noise on the VQE. J.C. performed the exact diagonalization, numerical optimization, and finite-size analysis. D.L. performed the circuit simulations for the performance of the VQE in the presence of noise. J.C., D.L., and A.B. wrote the manuscript. E.R. and M.M. supervised the project. All authors discussed the results and provided feedback for the manuscript.

\begin{acknowledgments}
J.C. and E.R. are supported by the grant PID2021-126273NB-I00 funded by MCIN/AEI/ 10.13039/501100011033 and by "ERDF A way of making Europe" and the Basque Government through Grant No. IT1470-22. This work was supported by the EU via QuantERA project T-NiSQ grant PCI2022-132984 funded by MCIN/AEI/10.13039/501100011033 and by the European Union ``NextGenerationEU''/PRTR. 

D.L. and M.M. gratefully acknowledge support by the German Ministry of Science and Education (BMBF) via the VDI within the projects MUNIQC-ATOMS and IQuAn, by the European Union via the ERC Starting Grant QNets, and by the Deutsche Forschungsgemeinschaft (DFG, German Research Foundation) under Germany’s Excellence Strategy ‘Cluster of Excellence Matter and Light for Quantum Computing (ML4Q) EXC 2004/1’ 390534769. This research is also part of the Munich Quantum Valley (K-8), which is supported by the Bavarian state government with funds from the Hightech Agenda Bayern Plus. Furthermore, the project leading to this publication has received funding from the European Union’s Horizon Europe research and innovation program under grant agreement No 101114305 (“MILLENION-SGA1” EU Project). 

A.B. acknowledges support from PID2021-127726NB-I00 (MCIU/AEI/FEDER, UE), from the Grant IFT Centro de Excelencia Severo Ochoa CEX2020-001007-S, funded by MCIN/AEI/10.13039/501100011033, and from the CSIC Research Platform on Quantum Technologies PTI-001. The project leading to this publication has received funding from the European Union’s Horizon Europe research and innovation program under grant agreement No 101114305 (“MILLENION-SGA1” EU Project).

The authors gratefully acknowledge the computing time provided to them at the NHR Center NHR4CES at RWTH Aachen University (Project No. p0020074). This is funded by the Federal Ministry of Education and Research and the state governments participating based on the resolutions of the GWK for national high-performance computing at universities.

This work has been financially supported by the Ministry for Digital Transformation and of Civil Service of the Spanish Government through the QUANTUM ENIA project call - Quantum Spain project, and by the European Union through the Recovery, Transformation, and Resilience Plan - NextGenerationEU within the framework of the Digital Spain 2026 Agenda.
\end{acknowledgments}

\appendix

\section{\bf Brief summary  of the $\mathbb{Z}_2$ LGT } \label{app:Z2}

The $\mathbb{Z}_2$ LGT has a simple structure: its gauge group is of order two and is cyclic. Its Hamiltonian contains only two competing terms which are tensor products of Pauli matrices. However, this does not make it a trivial model. In this Appendix, we review some details about its Hilbert space and phase diagram relevant to the analysis, and benchmark of the ground-state preparation algorithm.

The eigenstates of the $\mathbb{Z}_2$ LGT  Hamiltonian \eqref{eq:hamiltonian} are not analytically known in general. However, the physical eigenstates (Eq.~\eqref{eq:gauge_invariance}) of each of the individual terms ${H}_E$, ${H}_B$ can be easily derived. They provide a basis for the physical Hilbert space of the lattice. The states $\ket{\Omega_E}$ and $\ket{\Omega_B}$ defined in Eqs.~\eqref{eq:electric_gs} -- \eqref{eq:magnetic_gs} are the ground-states of ${H}_E$, ${H}_B$ respectively. The physical eigenstates of the electric field term consist of every possible product of plaquette operators (including the identity) acting over $\ket{\Omega_E}$, which can be understood as the combination of various electric-field strings along closed loops on the lattice
\begin{equation}
    \mathcal{B}_E = \{ \ket{\Omega_E}, {P}_{\boldsymbol{n}} \ket{\Omega_E}, {P}_{\boldsymbol{n}} {P}_{\boldsymbol{m}} \ket{\Omega_E},\dots\}.
\end{equation}
\noindent This set is used as a basis in the noiseless simulations in order not to consider gauge redundant basis states, reducing the overhead in memory. The physical eigenstates of the magnetic term ${H}_B$ are obtained by acting over $\ket{\Omega_B}$ with products of electric-field operators 
\begin{equation}
    \mathcal{B}_B = \{ \ket{\Omega_B}, {X}_{(\boldsymbol{n},i)}  \ket{\Omega_B} , {X}_{(\boldsymbol{n},i)} {X}_{(\boldsymbol{m},i)}  \ket{\Omega_B},\dots\}.
    \label{eq:plaquette_basis}
\end{equation}
which can be understood as introducing each of the possible different magnetic flux excitations. These bases are, in some sense, complementary because the ground state of ${H}_E$ (${H}_B$) can be expressed as a uniform superposition of every state in $\mathcal{B}_B$ ($\mathcal{B}_E$). 

\begin{figure}
    \centering
    \hspace{-0.5cm}
    \includegraphics[width=0.9\linewidth]{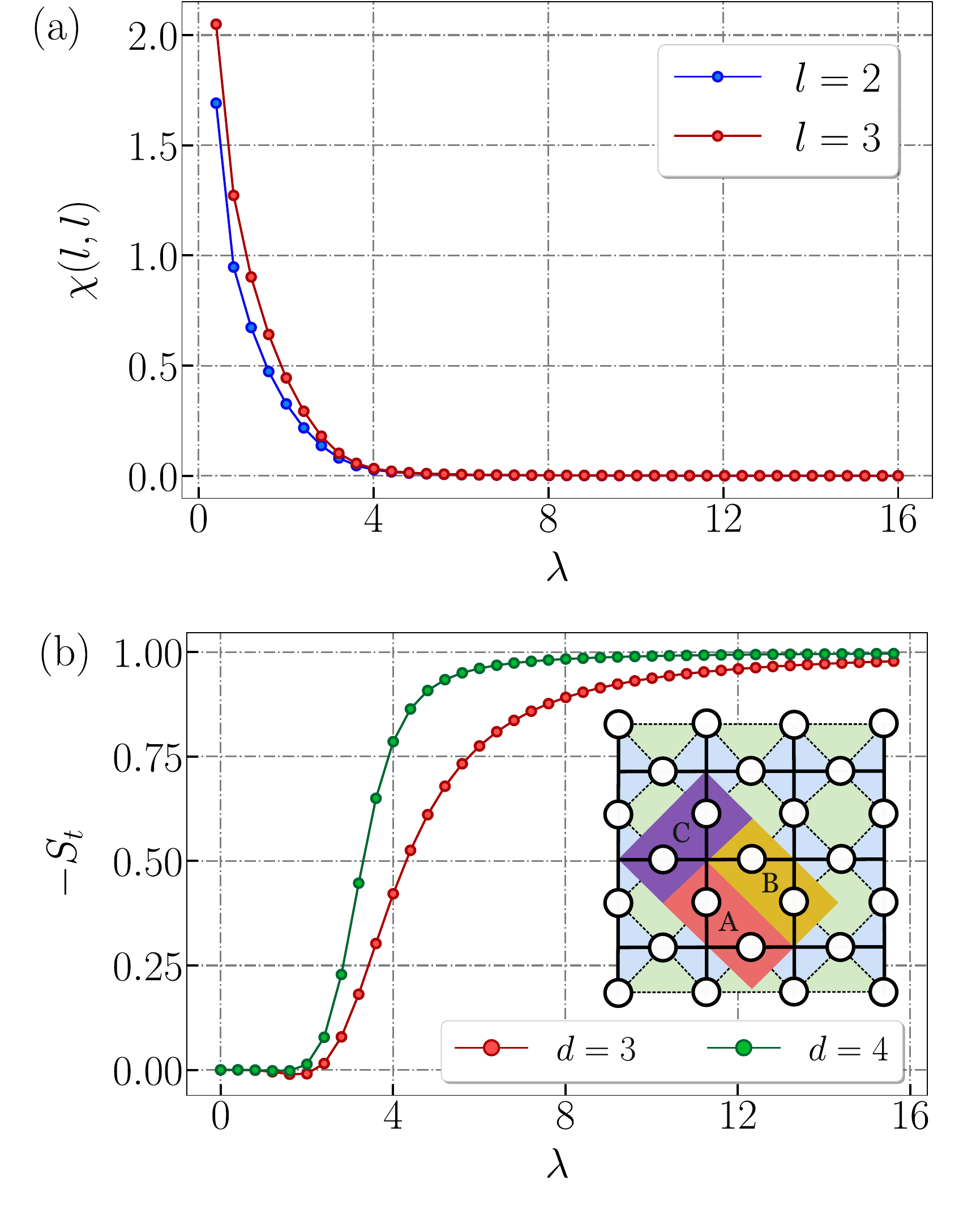}
    \caption{(a) Creutz ratio $\chi(l, l)$ in a $d = 5$ lattice for different loop sizes $l$. (b) Topological entanglement entropy of the subsystem that is colored in the inset. The subsystem is chosen to be as centered as possible.}
    \label{fig:noiseless_physical_quantities}
\end{figure}

\begin{figure*}[t]
    \includegraphics[width=1\linewidth]{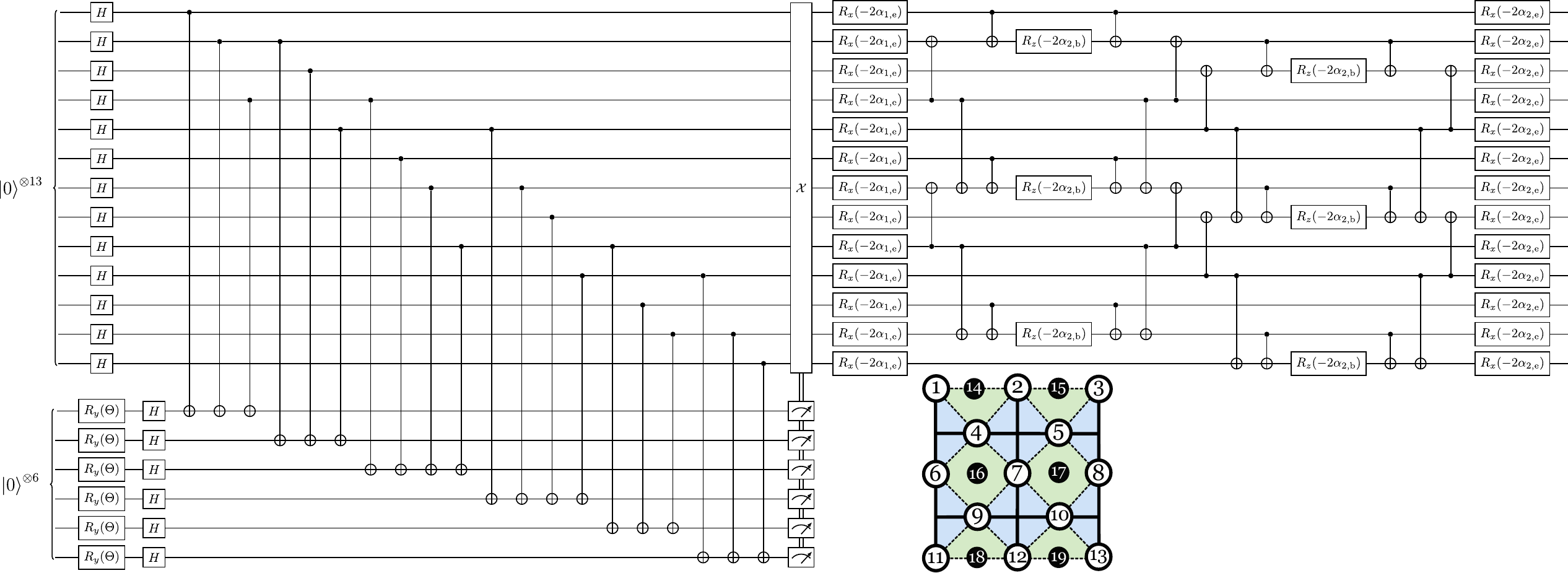}
    \caption{Quantum circuit implementation of two variational layers for the $d=3$ lattice, shown in the inset. The rotation angles of the ancilla qubits are $\Theta(\beta) = 2 \tan^{-1}(\tanh \beta)$. The operation $\mathcal{X}$ is a tensor product of $X$-operators acting on a subset of qubits determined from the ancilla measurement outcomes. Note that many operations in the circuit can in principle be parallelized, which is not shown here.}
    \label{fig:full_circuit}
\end{figure*}

The critical point of the $\mathbb{Z}_2$ LGT  $\lambda_c \simeq 3.044$ (Fig.~\ref{fig:title}(c)) coincides with that of the classical Ising model in 3 spatial dimensions. This is the case because the Hamiltonian \eqref{eq:hamiltonian} can be exactly mapped into that of the two-dimensional quantum Ising model in a transverse field through a Kramers-Wannier duality transformation \cite{kramers_wannier}, which is known to belong to the universality class of the classical $3D$-Ising Model. In Sec.~\ref{sec:noiseless_results} we provide an estimation of the $\beta, \nu$ critical exponents of this universality class coming from the proposed variational ansatz. The most precise value of these critical exponents that we have found in the literature is $\beta = 0.3265(3)$, $\nu = 0.6301(4)$ \cite{ising_critical_exponents}. On the other hand, the nature of the phases is very different in both models. Elitzur's theorem \cite{elitzur} forbids the existence of local order parameters that can label the different phases of the $\mathbb{Z}_2$ LGT. Instead, a couple of non-local order parameters can achieve this goal. The first quantity resembling the behavior of an order parameter is the expectation value of Wilson loop operators
\begin{equation}
    {W}_{\mathcal{C}} = \prod_{(\boldsymbol{n}, i) \in \mathcal{C}} {Z}_{(\boldsymbol{n}, i)} = \prod_{\boldsymbol{n} \in \mathcal{A}_{\mathcal{C}}} {P}_{\boldsymbol{n}}.
\end{equation}
They consist of products of ${Z}_{\boldsymbol{n}, i}$ operators around a closed loop $\mathcal{C}$ on the lattice (Fig.~\ref{fig:title}(a)), and can be expressed as the product of every plaquette contained in the interior region of the loop $\mathcal{A}_{\mathcal{C}}$. Wilson loops are used in the LGT  formalism to detect the presence of confinement. The expectation value of this operator does not behave as a usual order parameter\footnote{Vanishing in one phase, taking non-zero values in the other, and scaling law on the lattice size}. Instead, its scaling with the size of the considered closed loop is different in each phase
\begin{gather*}
    \text{Confined phase $\lambda<\lambda_c$:} \hspace{0.82cm} \langle {W}_{\mathcal{C}} \rangle \sim {\rm e}^{-\chi A_\mathcal{C}} \\
    \text{Deconfined phase $\lambda>\lambda_c$:} \hspace{0.5cm} \langle {W}_{\mathcal{C}} \rangle \sim {\rm e}^{-\Gamma \ell_\mathcal{C}}.
\end{gather*}
\noindent Here, $A_{\mathcal{C}}$, $\ell_{\mathcal{C}}$ the area and perimeter of the considered loop, respectively. This behavior can be used to define an ordinary order parameter. The coefficient of exponential decay in the confined phase $\chi$ can be extracted through the so-called Creutz ratio \cite{creutz_ratio}, namely
\begin{equation}
    \chi(l, l) = -\log \frac{\langle {W}_{(l, l)} \rangle \langle {W}_{(l-1, l-1)} \rangle}{\langle {W}_{(l-1, l)} \rangle \langle {W}_{(l, l-1)} \rangle},
    \label{eq:creutz}
\end{equation}
\noindent 
where we have defined ${W}_{(l, m)}$ as a rectangular Wilson Loop of side length $l \times m$ in lattice units. The Creutz ratio behaves as a usual order parameter \cite{creutz_scaling}: returning the coefficient of exponential decay $\chi>0$ in the confined phase and vanishing $\chi=0$ in the deconfined one. $\chi$ is related to the string tension in the confining phase, which can be proven to be proportional to $\lambda^{-1}$, and depends on the size of the  Wilson loops \cite{creutz_ratio, creutz_scaling}. Figure~\ref{fig:noiseless_physical_quantities}(a) shows the data for the Creutz ratio from the DVA  in the largest considered lattice for different values of $l$.

Another order parameter that we consider is the expectation value of the dual magnetization $\langle {M} \rangle$. The operator ${M}$ is a tensor product of ${X}_{(\boldsymbol{n}, i)}$ operators along a path $\partial \mathcal{C}_n$ defined on the dual lattice, which extends from its boundary to one of the links contained in plaquette ${P}_{\boldsymbol{n}}$, as depicted in the scheme of  Fig.~\ref{fig:title}(a).
\begin{equation}
    {M}_n = \prod_{(\boldsymbol{n}, i) \in \partial \mathcal{C}_n} {X}_{(\boldsymbol{n}, i)}.
    \label{eq:dual_magnetization_app}
\end{equation}
We note that the dual magnetization maps onto the spin operator of the dual transverse-field Ising model through the Kramers-Wanniers duality transformation \cite{kramers_wannier}, and thus serves as an order parameter. In the LGT  formalism, this kind of operator is usually known as the magnetic t'Hooft string \cite{fradkin_2013}.

We also note that the deconfined phase of the $\mathbb{Z}_2$ LGT  is a topological phase with long-range entanglement. These are a special kind of quantum (zero-temperature) phase in which the ground state of the system becomes degenerate in the thermodynamic limit, such that this degeneracy can not be lifted through local perturbations, and depends on the non-trivial homology that underlies the lattice model. For instance, if the gauge theory is embedded in a torus, the non-contractible paths around and across the hole lead to a robust ground-state degeneracy. For the planar version with specific boundary conditions such as the one that we consider, one can also define non-contractible paths connecting the opposite boundaries, leading also to a topological degeneracy \cite{Dennis_2002, low_distance_surface_codes}. Our choice of the boundary conditions leading to this degeneracy without the need for periodic boundary conditions is not by any means unique \cite{low_distance_surface_codes}, as an example, one could also choose boundary conditions resembling the rotated surface code \cite{low_distance_surface_codes}.

In topological-ordered phases, an additional manifestation of the non-trivial topology appears in the Rényi entropy. Considering the reduced density matrix of subsystem $A$, which is obtained by tracing the ground-state density matrix over the complement $\bar{A}$ $\rho_A={\Tr}_{\bar{A}}\{\ket{\rm gs}\bra{\rm gs}\}$, the $\alpha$-th order Rényi entropy\footnote{The Von-Neumann entropy is considered here as the Renyi entropy with $\alpha \to 1$ for brevity.} is defined as 
\begin{equation}
    S_A^{(\alpha)} = \frac{1}{1 - \alpha} \log_2 \Tr \left[ \rho_A^\alpha \right]
\end{equation} 
It can be proven that this entropy has a universal contribution \cite{renyi_universaility, entanglement_kitaev, entanglement_enrique}, known as the topological entanglement entropy
\begin{equation}
    S_{t} = S^{(\alpha)}_A + S^{(\alpha)}_B + S^{(\alpha)}_C - S^{(\alpha)}_{AB} - S^{(\alpha)}_{AC} - S^{(\alpha)}_{BC} + S^{(\alpha)}_{ABC},
    \label{eq:top_entanglement_entropy}
\end{equation}
where the subsystems $A$, $B$, and $C$ must be defined such that they all share a boundary. The value of the topological entanglement entropy for the $\mathbb{Z}_2$ LGT  can be computed analytically in the extreme cases $\lambda \to \{0, \infty\}$, where it takes values $S_t \in \{0, -1\}$ respectively \cite{burrelo, entanglement_computation}. For generic $\lambda$, numerical methods must be used to show that $S_t=1 $ in the whole deconfined phase $\lambda>\lambda_{\rm c}$. Remarkably, a possible way to infer the entanglement entropy from a QS is to measure the second-order Rènyi entropy for each subsystem through the randomized measurement scheme \cite{renyi_random_measurement, randomized_measurement_ltg, randomized_measurement_toolbox}, which provides access to the trace of the $\alpha$-th power of the reduced density matrix of any subsystem through repeated measurements in random bases. In Fig.~\ref{fig:noiseless_physical_quantities}(b) we show the topological entropy of the dissipative variational state. The introduction of the dissipative layer enables the ansatz to transition from a state with short-range entanglement in the confined phase to a highly entangled state in the topological phase while maintaining a good approximation in between.

\section{\bf Quantum circuit implementation} \label{app:circuit}

Figure~\ref{fig:full_circuit} shows a quantum circuit implementing two variational layers of our proposed DVA  for a $d=3$ lattice. The inset in the figure shows a potential embedding of the $\mathbb{Z}_2$ LTG on a quantum processor. Qubits are labeled from top to bottom. Note that the train of CNOT gates which preludes the measurement of the ancilla qubits can be executed in four steps of parallel executions of two-qubit gates, as discussed in Sec.~\ref{sec:circuit_implementation}. We display the CNOT gates acting in series for clarity. Also, the latter part of the circuit can be parallelized, as discussed in Ref.~\cite{burrelo}.

\section{\bf Additional data on the DVA performance} \label{app:diss_vs_unitary}

In this Appendix, we provide more details on the quality of the variational states prepared with our dissipative variational ansatz \eqref{eq:ansatz} and both unitary HVA ans\"atze \eqref{eq:HVA} and \eqref{eq:HVA_B}. It is important to remark that we have used the optimization strategy outlined in Appendix~\ref{app:optimization_algorithm} to compute the optimal variational parameters also with the unitary ansatz, and we have found that the unitary HVA ansatz is more expressible than the results in \cite{burrelo} show. Still, the DVA provides an advantage in terms of a lower number of layers required to achieve good precision, especially around the critical region of the model. The cost of this improvement is the introduction of ancilla qubits (up to one per plaquette in the lattice) but this overhead is not prohibitive in realistic quantum devices as described in Sec.~\ref{sec:circuit_implementation}.

\begin{figure}
    \centering
    \includegraphics[width=0.9\linewidth]{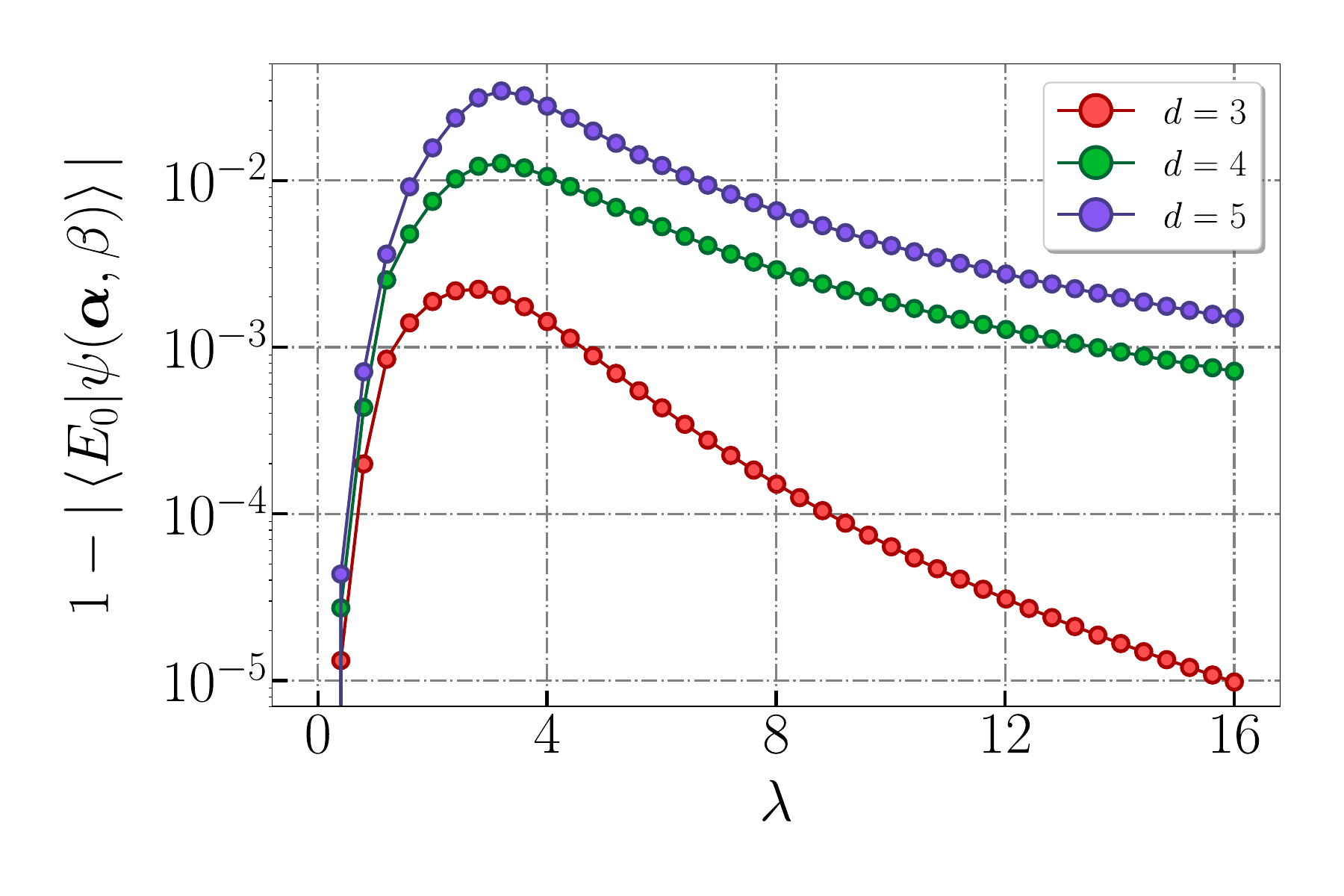}
    \caption{Infidelity of variational states produced with the dissipative variational ansatz with $\ell=2$ layers w.r.t. the true ground states at different values of $\lambda$ for increasing lattice size. The profile of these curves is very similar to that of the relative energy difference shown in Fig.~\ref{fig:noiseless_ansatz_performance} in the main text.}
    \label{fig:infidelity}
\end{figure}

In Fig.~\ref{fig:infidelity} we show the infidelity of the variational states prepared with $\ell=2$ layers of our dissipative ansatz w.r.t. the true ground states of the $\mathbb{Z}_2$ LGT up to a lattice size $d=5$ in the noiseless regime. Note that the curves look very similar to the ones in Fig.~\ref{fig:noiseless_ansatz_performance}, which implies that the minimum energy is equivalent to the maximum achievable fidelity. This check is required to ensure the correctness of the variational method since in a topological model the lower part of the Hamiltonian spectrum is degenerate.

\begin{figure}
    \centering
    \includegraphics[width=0.9\linewidth]{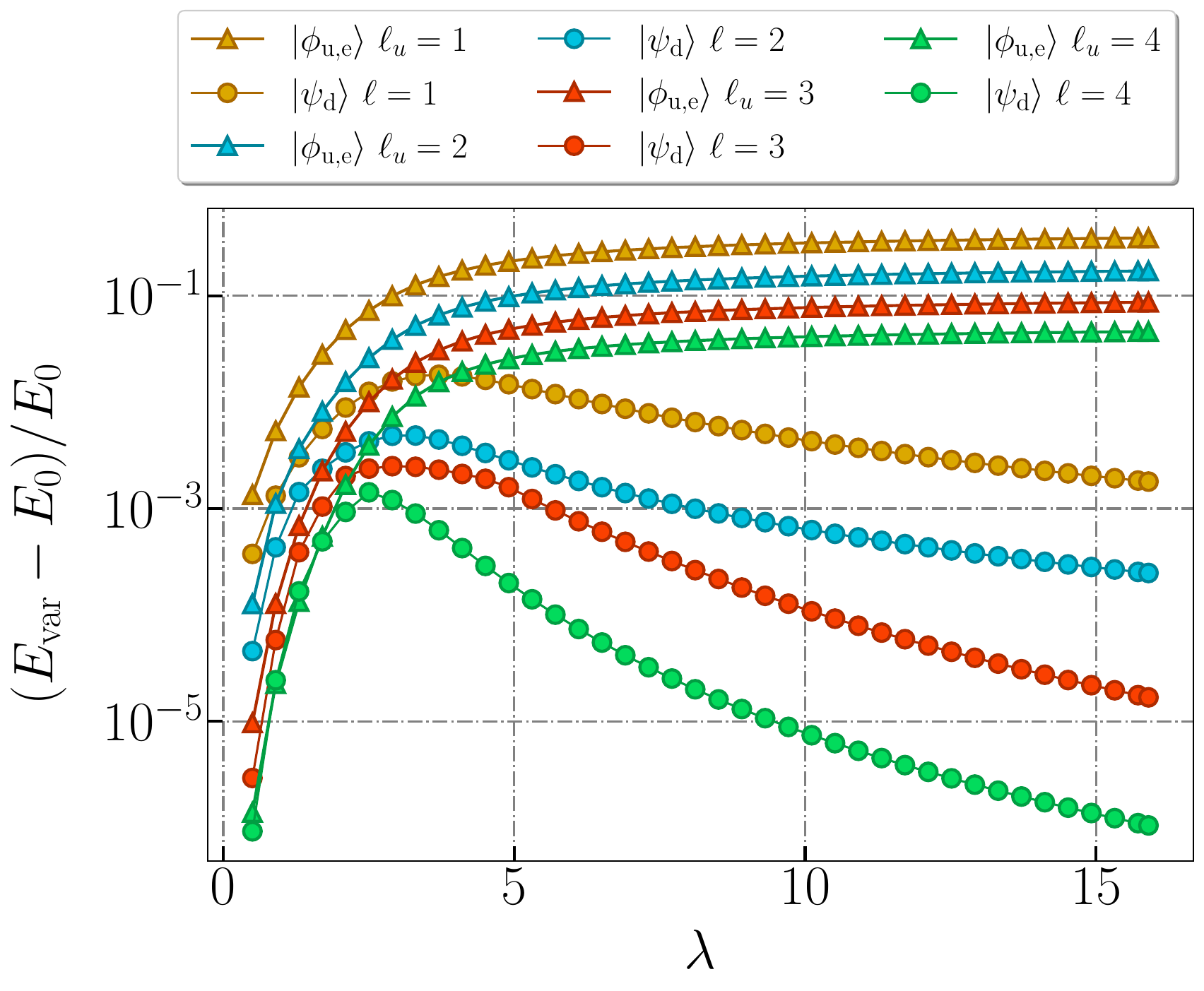}
    \caption{Relative difference between the exact ground-state energy and the energy achieved by the dissipative and electric unitary ansätze for increasing number of layers and different values of $\lambda$ in the $d = 4$ lattice.}
    \label{fig:unitary_dissipative_energy_comparison}
\end{figure}

Figure~\ref{fig:unitary_dissipative_energy_comparison} shows the relative energy differences between the variational energy and the true ground state energy for our dissipative proposal as well as for the electric unitary ansatz \eqref{eq:HVA} in the $d = 4$ lattice with an increasing number of layers. As expected, the dissipative ansatz consistently achieves lower energies for shallow depths $\ell < d$ at every value of $\lambda$. For $\ell \geq d$ the unitary ansatz becomes expressive enough to achieve energies comparable to the dissipative ones in the electric phase of the model. In this regime, the true ground state of the model is closer to the initial product state $\ket{\Omega_E}$ and a set of unitary layers of size comparable to the side length of the lattice can efficiently generate it. When both ans\"{a}tze are expressive enough, there are values of $\lambda$ inside the electric phase of the LTG for which the unitary ansatz achieves slightly lower energies. We expect both ansätze to be equivalent in the infinite-depth regime. In the critical region of the model, a considerable number of layers must be introduced for this to happen. 

\begin{figure*}[t]
    \includegraphics[width=0.8\linewidth]{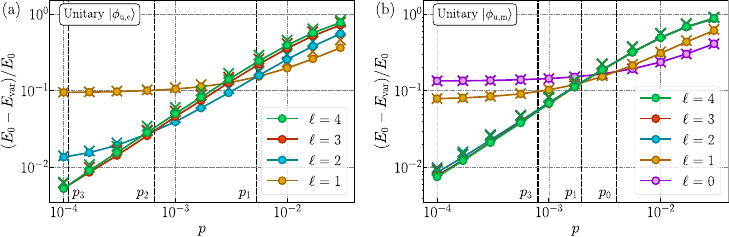}
    \caption{Precision of the estimated ground-state energy for the $d=3$ lattice, at  $\lambda = 3.0$. Variational states are prepared by applying $\ell$ layers of the unitary ansatz starting from (a) the trivial reference state $\ket{\Omega_E}$ and (b) the surface code reference state $\ket{\Omega_B}$. The states are prepared in the presence of circuit-level noise of strength $p$. Preparing just $\ket{\Omega_B}$ corresponds to $\ell=0$ variational layers in panel (b).}
    \label{fig:noisy_GS_energy_uni}
\end{figure*}

In Fig.~\ref{fig:noisy_GS_energy_uni} we show the ground-state energy precision achieved by the unitary ans\"atze \eqref{eq:HVA} and \eqref{eq:HVA_B} in the presence of circuit-level noise in the $d = 3$ lattice. This is data used for the comparison between dissipative and unitary state preparation in Fig.~\ref{fig:noisy_comparison_diss_uni} in the main text, and displays the consecutive variational thresholds for the error rates which are required to increase the layer number in each ansatz. These thresholds are expected to decrease when considering larger lattices due to a higher number of CNOT gates present in the variational state preparation circuits.

\section{\bf Scaling and correction exponents} \label{app:curve_collapse}

In a quantum phase transition at zero temperature, the characteristic energy scale of the system is the gap $\Delta$ between its ground and first energy state. In a second-order continuous phase transition like the one under consideration, this quantity scales as
\begin{equation}
    \Delta \sim \left| \lambda - \lambda_c \right|^{\nu z}.
\end{equation}
Here, $\nu$ and $z$ are the correlation length and dynamical critical exponents, respectively, and $\lambda_c$ is the critical coupling strength in Hamiltonian \eqref{eq:hamiltonian}. We note that this is only the leading scaling and that higher-order corrections will appear explicitly later that can play a role in small systems. The mapping relating the $\mathbb{Z}_2$ LTG and the quantum Ising model in (2+1)-dimensions indicates that $z=1$ \cite{sachdev2007}. This vanishing of the gap causes the correlation length $\xi$ to diverge as
\begin{equation}
    \xi \sim \xi_0 |\lambda - \lambda_c|^{-\nu}.
    \label{eq:correlation_length}
\end{equation}
Similarly, the order parameter, denoted as $\langle M \rangle$ anticipating the use of the dual magnetization as an order parameter, around the transition scales as
\begin{equation}
    \langle M \rangle \sim M_0 |\lambda - \lambda_c|^{\beta}.
    \label{eq:magnetization_scaling}
\end{equation}
In finite lattices of lateral size $L = d$, the correlation length cannot take infinite values, and the characteristic length of the system in the vicinity of the phase transition thus becomes the size of the system $L$. One then makes the following identification
\begin{equation}
    |\lambda - \lambda_c| \sim \xi^{-1/\nu} \sim L^{-1/\nu}.
    \label{eq:lambda_c_scaling_simple}
\end{equation}

\begin{figure}[!ht]
    \centering
    \includegraphics[width=0.9\linewidth]{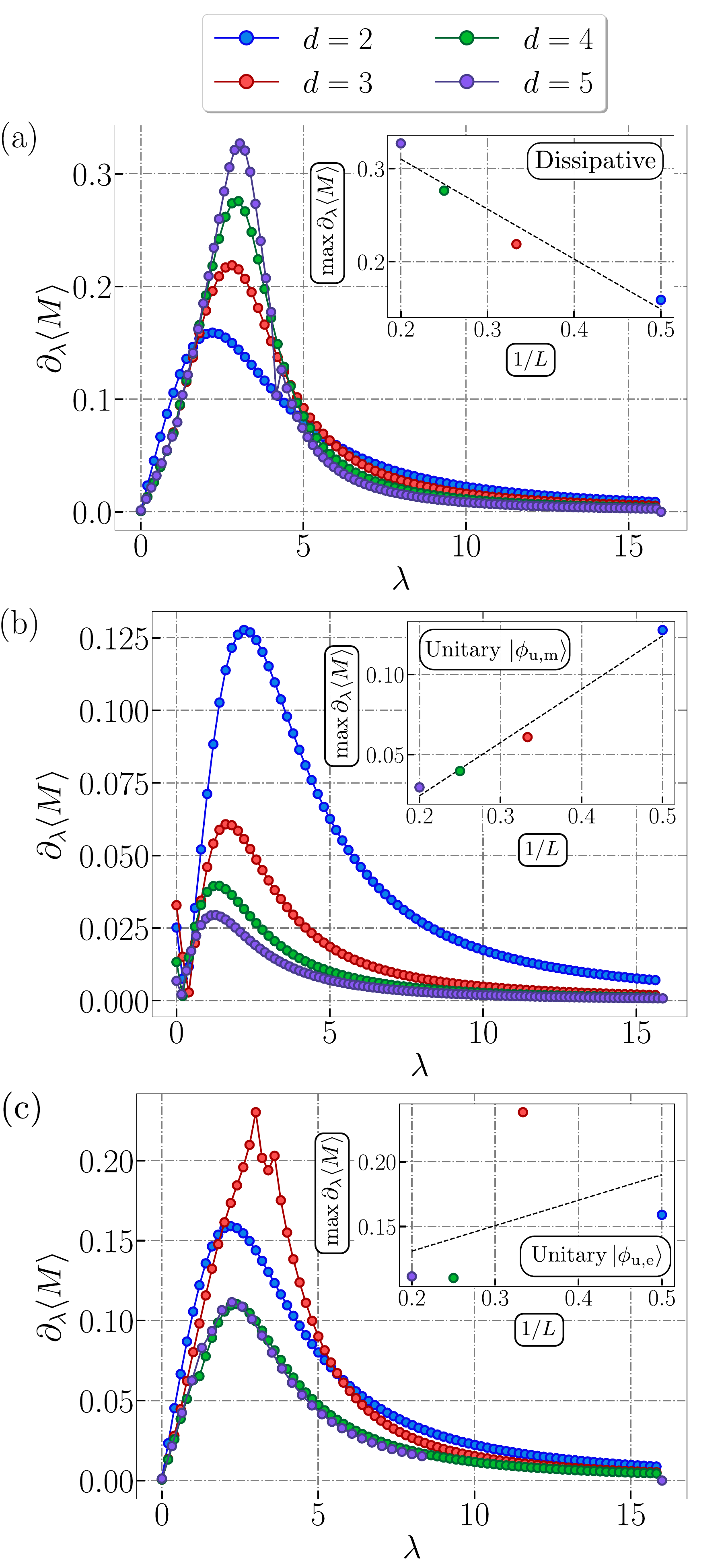}
    \caption{Dual magnetic susceptibilities for the (a) dissipative ansatz, (b) the unitary magnetic ansatz with reference state $\ket{\Omega_B}$ and (c) the unitary electric ansatz with reference state $\ket{\Omega_E}$ as a function of the microscopic coupling $\lambda$, and considering various distances $L=d$ with the corresponding increase in the lattice size. In the insets, we display the maximum value of the susceptibility as a function of $1/L$. The slopes would give an estimate of $\nu$ in the thermodynamic limit, discarding logarithmic corrections. The dashed lines are only a guide to the eye to emphasize the opposite trends. The dissipative ansatz captures the expected behavior.}
    \label{fig:susceptibilities}
\end{figure}

Now including the aforementioned higher-order corrections explicitly, substituting \eqref{eq:lambda_c_scaling_simple} into \eqref{eq:magnetization_scaling}, one arrives at the finite size scaling ansatz for the order parameter
\begin{equation}
    \langle M \rangle = a L^{-\beta/\nu}(1 + bL^{-\theta/\nu} + \dots),
    \label{eq:magnetization_scaling_corrections}
\end{equation}
where $\theta$ is the so-called first correction exponent, which has the value $\theta = 0.52$ in the 3D-Ising universality class \cite{LUNDOW201840}, and $a, b$ non-universal constants. This scaling law is valid as long as the variable
\begin{equation}
    x = (\lambda - \lambda_c) L^{1/\nu}
\end{equation}
is kept fixed. This implies that the critical behavior in a finite lattice of size $d$ should be observed around the following values of the coupling $\lambda$
\begin{equation}
    \lambda_c(d) = \lambda_c(\infty) + a' L^{-1/\nu}(1 + b' L^{-\theta/\nu})
    \label{eq:lambda_c_scaling}
\end{equation}

\begin{figure}
    \centering
    \includegraphics[width=0.9\linewidth]{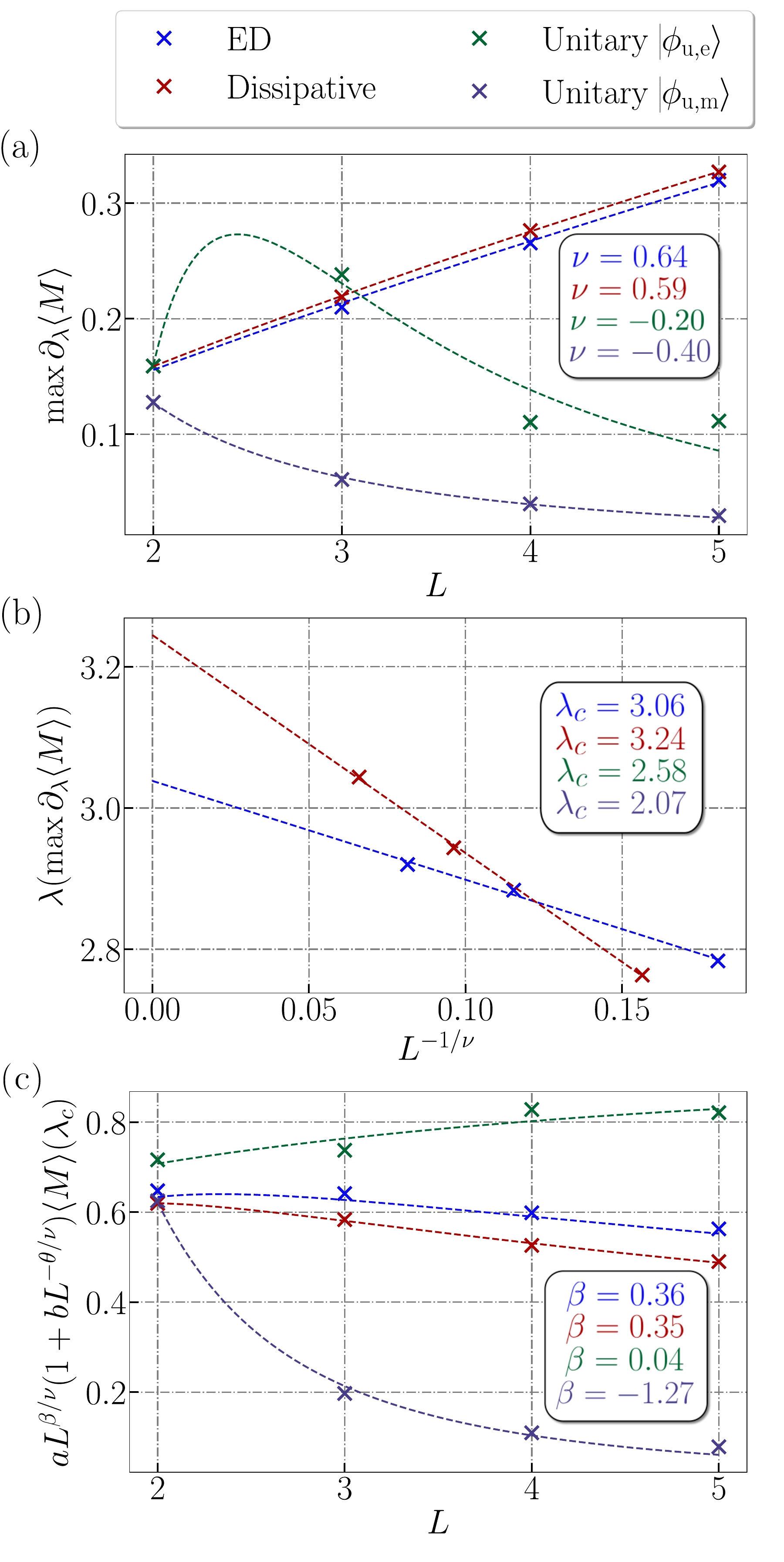}
    \caption{Finite size scaling using data from ED and the different variational states with the same depth corresponding to $\ell = 2$ in the dissipative ansatz. We use the exact value for the correction exponent $\theta$ for all fits. (a) Fits for Eq.~\eqref{eq:max_mag_scaling}, providing the estimation of $\nu$. (b) Fits for Eq.~\eqref{eq:lambda_c_scaling}, providing the estimation of $\lambda_c$. We omit the curves of the unitary ans\"atze because the negativity of $\nu$ leads to an unreadable plot. (c) Fits for Eq.~\eqref{eq:magnetization_scaling_corrections}, providing the estimation of $\beta$.}
    \label{fig:fssa_fits}
\end{figure}

These expressions can be fitted to the dual magnetization data that can be extracted from the various variational ans\"atze or the exact diagonalization to extract the critical exponents of the $\mathbb{Z}_2$ LTG. Since we only have access to small lattice sizes, in which finite-size effects can be large, we do not expect to be able to attain extremely precise estimations of the critical exponents in comparison to other numerical approaches of the corresponding dual Ising models. On the other hand, the variational approach deals with the real model and could be used to explore other situations of interest, including other gauge sectors, and eventually be generalized to encompass real-time dynamics and finite-density matter. In this Appendix,  we use the critical exponent estimates to quantify the advantage of the DVA  against the unitary one in the study of physical observables. We compare the performance of the variational ans\"atze against the exact diagonalization to distinguish finite-size effects.

We first use the fact that the maximum value of the derivative of the dual magnetization (dual magnetic susceptibility) scales with the lattice size as \cite{PhysRevB.44.5081}
\begin{equation}
    \max \left( \frac{\partial \langle M \rangle}{\partial \lambda} \right) = a'' L^{1/\nu}(1 + b''s L^{-\theta/\nu}).
    \label{eq:max_mag_scaling}
\end{equation}
This allows us to extract the value of the $\nu$ exponent, using the exact value $\theta$ as a given input. The unitary HVA predicts that the peak of the magnetic susceptibility decreases with increasing system size(Fig.~\ref{fig:susceptibilities}), leading to a negative value of $\nu$. This contradicts what is expected in general when performing a finite-size scaling analysis. The value of $\lambda_c$ is extracted by fitting the position of the maximum derivative of $\partial \langle M \rangle/ \partial\lambda$ (see Fig.~\ref{fig:susceptibilities}) to Eq.~\eqref{eq:lambda_c_scaling}. Similarly, $\beta$ is estimated by fitting the dual magnetization data to \eqref{eq:magnetization_scaling_corrections}. Figure~\ref{fig:fssa_fits} shows the results of this analysis, including also the exact diagonalization data. The maximum value of the magnetization derivative predicted by the unitary ansatz becomes lower causing the fit to return a negative value of $\nu$. The values predicted by the DVA are reasonably close to the exact value, considering the small lattices that are used.

\section{\bf Details on numerical methods} \label{app:optimization_algorithm}

\begin{figure}[t]
    \centering
    \includegraphics[width=0.8\linewidth]{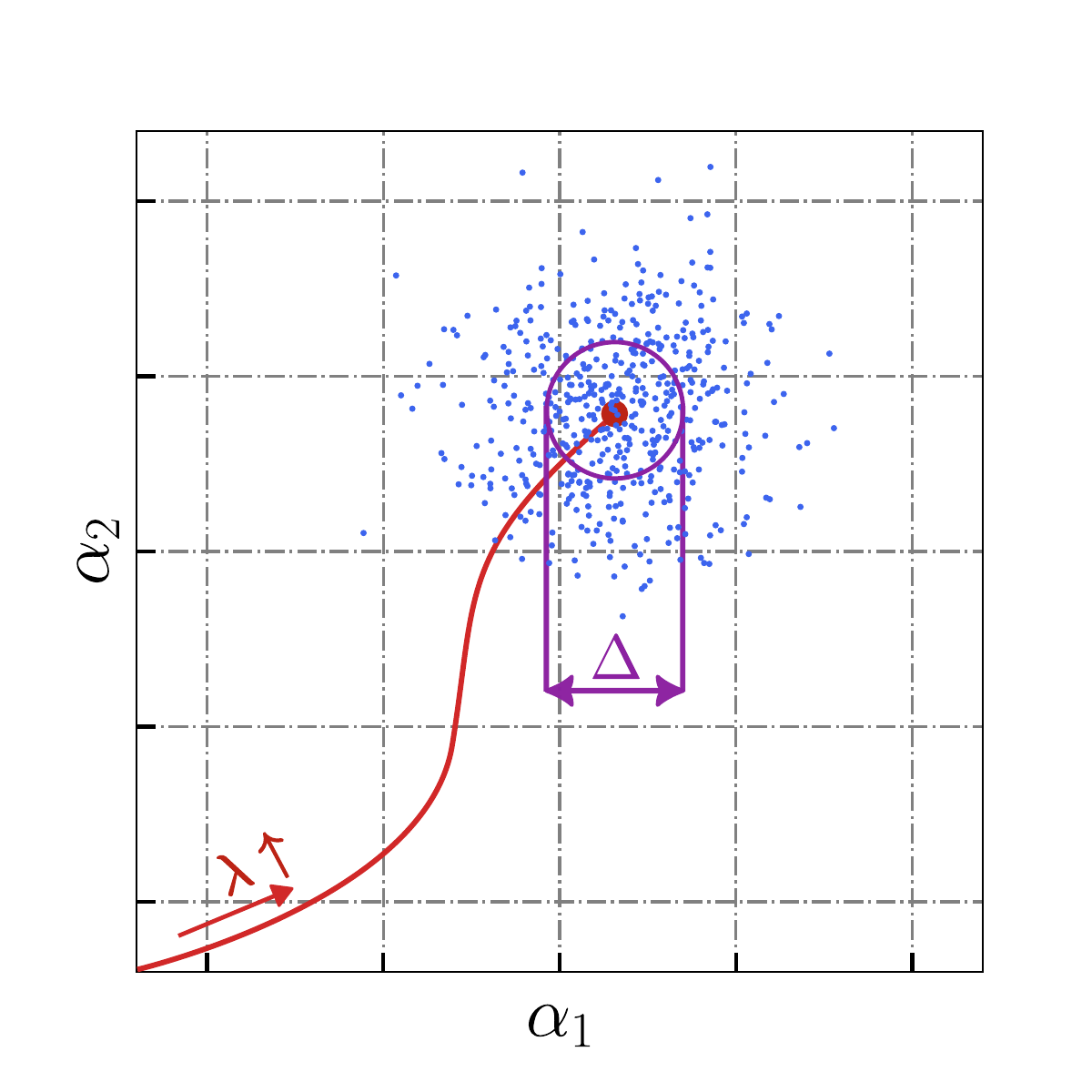}
    \caption{Sketch of the optimization strategy (Algorithm~\ref{alg:optimizationStrategy}). Blue dots represent the set of seeds for the next batch of gradient descent optimization algorithms. They are sampled from a $2L$-dimensional Gaussian distribution with variance matrix $\sigma = \mathrm{diag}(\Delta, \Delta \dots \Delta)$. The positions of the minima found for each value of $\lambda$ describe a trajectory in parameter space that is not necessarily continuous.}
    \label{fig:sketch-optimization}
\end{figure}

\paragraph{Optimization algorithm}
A critical ingredient ensuring the efficiency of VQEs is the initialization of the variational parameters. The energy expectation value of almost every variational ansatz is a highly non-convex function in the space of parameters, which makes the optimization process difficult. The energy minimization in VQEs is usually performed using classical gradient descent algorithms. These optimizations converge quickly to a solution, but they can get trapped in local minima, thus not providing the best possible variational parameters. Even though global optimization algorithms exist, they usually converge slowly and neither provide guarantees of success. We have opted for designing an initialization strategy for the variational parameters that increase the probability of finding the global minimum. It is based on the assumption that the value of the optimal variational parameters must change only slightly when $\lambda$ is changed by a small $\delta \lambda$. When $\lambda = 0$ the reference state is already the true ground-state and the optimal value of all the variational parameters is known to be identically zero. We then propose the strategy to find the optimal variational parameters for increasing values of $\lambda$ outlined in Algorithm~\ref{alg:optimizationStrategy}. Figure~\ref{fig:sketch-optimization} shows a schematic of the optimization strategy.

We empirically have found that this strategy can extract the potential of the proposed ansatz, observing little variation among different runs and consistently providing better results than random initialization. It is worth mentioning that we do not take advantage of the transferability of the variational parameters among lattice sizes that we certainly observe, in correspondence with the results in \cite{burrelo}, because our strategy is good enough. In any case, this principle could be easily included if needed.  

For the simulation of the noiseless ground-state preparation process, we have performed a state-vector simulation implemented in a Python environment using the Scipy ecosystem \cite{numpy, scipy}. For the gradient descent subroutine, we have used the implementation of the L-BFGS-B \cite{gradient_descent} algorithm provided by this framework. This optimization algorithm can accommodate the constraints on the possible values of the variational parameters $\beta_1 \in [0, 1]$, $\alpha_k, \beta_k \in [0, 2\pi]$. We distribute each of the gradient descent instances among the different cores of the processor we use, it is interesting to note that this parallelization scheme could be also used in a future implementation of this variational algorithm on a family of real quantum devices that can be controlled by a single classical interface. We use the following input parameters in our simulations of the ground-state preparation process $N_\lambda = 800$, $\lambda_{\mathrm{max}} = 16$, $N_s = 336$, $\Delta = 0.1$.

\paragraph{Simulation of noisy circuits}

To simulate noise during the variational state preparation, we sample $10^5$ erroneous realizations of the preparation circuit, where random Pauli errors are sampled after every circuit operation, as described in Sec.~\ref{sec:results_noise} of the main text. We introduce one ancilla qubit per plaquette in the noisy simulations. We use each circuit realization to prepare a variational state and measure all qubits of this state 100 times in the $Z$-basis. This is done because measurements of a given state are computationally much less demanding than the preparation of a variational state. We repeat the whole procedure for final measurements on the $X$ basis.

\begin{algorithm}[H]
    \caption{Optimization strategy}\label{alg:optimizationStrategy}
    \begin{algorithmic}
        \Require $N_\lambda \in \mathbb{N}$ \Comment{Number of lambda values to consider}
        \Require $\lambda_{\mathrm{max}} > 0$ \Comment{Maximum value of $\lambda$}
        \Require $N_{s} \in \mathbb{N}$ \Comment{Optimization trials per $\lambda$}
        \Require $\Delta \in \mathbb{R}_{+}$ \Comment{Sample variance}
        \State $\mathrm{VP}_{(\boldsymbol{n}_\lambda \times 2L)} \gets \mathrm{diag}(0, 0, \dots, 0)$ \Comment{Matrix storing parameters}
        \State $n \gets 2$
        \For{$\lambda \in \lambda_2, \lambda_3 \dots \lambda_{N_\lambda}$}
            \State $\mathrm{\vb*{\alpha}_{\mathrm{prev}}} \gets \mathrm{row}_{n - 1}(\mathrm{VP})$ \Comment{Previous variational params.}
            \State Sample a set $\left\{ \vb*{\alpha}_k \right\}_{k = 0}^{N_\mathrm{s}}$ of parameter vectors from a multivariate gaussian with $\vb*{\mu} = \vb*{\alpha}_{\mathrm{prev}}$ and $\sigma_{(2L \times 2L)} = \mathrm{diag}(\Delta, \dots \Delta)$. Use them to initialize $N_s$ instances of a gradient descent algorithm. Use periodic boundary conditions.
            \State Let $\vb*{\alpha}_{\mathrm{opt}}$ be the vector of parameters leading to the variational state with minimal energy found after executing all gradient descent instances.
            \State $\mathrm{row}_{n}(\mathrm{VP}) \gets \vb*{\alpha}_{\mathrm{opt}}$
            \State $n \gets n + 1$
        \EndFor
        \State \textbf{return} $\mathrm{VP}$
    \end{algorithmic}
\end{algorithm}

\section{\bf Parameter-shift rule for gradient descent} \label{app:parameter-shift-rule}

Classical optimization algorithms perform many calls to the cost function to compute gradients. This is an issue in the context of VQEs, where the cost function is the energy expectation value of the variational state, whose measurement requires many repetitions of the ground-state preparation circuit. The parameter-shift rule \cite{parameter_shift_og, parameter_shift_good} is a technique for exactly computing derivatives of expectation values to the parameters of variational states through the measurement of a reduced set of expected values. It is a key tool that considerably reduces the number of circuit repetitions needed to perform energy minimization in the quantum device. The parameter shift rule is usually presented in the context of unitary ground-state preparation, but unitarity is not a requirement. The true prerequisite enabling the parameter-shift rule is that all the parameterized operations in the ground-state preparation circuit can be expressed as the exponential of some operator. To present the extension of the parameter shift rule to the variational ansatz presented in this work two cases must be distinguished, the gradient component associated with the parameter in the non-unitary operation $\beta_1$ and the rest. The first thing to notice in any case is that it is possible to write the expectation value of an observable ${O}$ over a variational state as follows
\begin{equation}
   \frac{\partial \langle {O} \rangle}{\partial \alpha_k} = 2 \Re  \left[ \bra{\psi(\vb*{\alpha}, \beta)} {O} \; \frac{\partial \ket{\psi(\vb*{\alpha}, \beta)}}{\partial \alpha_k} \right].
   \label{eq:observable_derivative}
\end{equation}
\noindent The derivative of the variational ground-state can be prepared with a circuit very similar to that preparing the state itself, which allows measuring the derivative of the observable directly on the quantum computer. Notice that 
\begin{equation}
    \begin{split}
        \frac{\partial \ket{\psi(\vb*{\alpha}, \beta)}}{\partial \beta} = & \sum_{n=0}^{N_p} U(\vb*{\alpha}) {P}_n \, \frac{e^{\beta {H}_B}}{(\cosh 2\beta_1)^{N_p/2}} \ket{\Omega_E} \\[5pt]
        & - \tanh \beta \ket{\psi(\vb*{\alpha}, \beta)}
    \end{split}
    \label{eq:states_derivatives_1}
\end{equation}
\begin{equation}
    \begin{split}
        \frac{\partial \ket{\psi(\vb*{\alpha}, \beta)}}{\partial \alpha_{k, \mathrm{b}}} = \left[ \sum_{n = 0}^{N_p} \right. & \mathcal{O}_{(k+1, L)}(\vb*{\alpha}) e^{i \alpha_k {H}_E} (i {P}_n) \\
        & \left. \times e^{i \gamma_k {H}_B} \tilde{\mathcal{O}}_{(1, k-1)} (\tilde{\vb*{\alpha}}, \beta) \right] \ket{\Omega_E}
    \end{split}
    \label{eq:states_derivatives_2}
\end{equation}
\begin{equation}
    \begin{split}
        \frac{\partial \ket{\psi(\vb*{\alpha}, \beta)}}{\partial \alpha_{k, \mathrm{e}}} = \left[ \sum_{n, i} \right. & \mathcal{O}_{(k+1, L)}(\vb*{\alpha}) (i {X}_{(\boldsymbol{n}, i)}) e^{i \alpha_k {H}_E} \\
        & \left. \times e^{i \gamma_k {H}_B} \tilde{\mathcal{O}}_{(1, k-1)} (\vb*{\alpha}, \beta) \right] \ket{\Omega_E}
    \end{split}
    \label{eq:states_derivatives_3}
\end{equation}
\noindent With $U(\vb*{\alpha})$ the set of unitary operations contained in the variational ansatz and $\tilde{\mathcal{O}}_{(1, k-1)} (\tilde{\vb*{\alpha}}, \tilde{\vb*{\beta}})$ the set of operations in layers $1$ to $k$. Expressions \eqref{eq:observable_derivative}-\eqref{eq:states_derivatives_3} imply that the gradient of the expectation value of an observable computed over the proposed variational state can be exactly computed from $4N_p + 2N$ expectation values. That is, two per term contained in the sums above. These expectation values are measured over states that can be prepared with circuits almost identical to those generating the ground state. Just a single extra unitary is introduced after the $k$-th layer of the ansatz. The expression for the derivatives to the parameters associated with the unitary operations is identical to what can be found in the literature \cite{variational_filters}. The derivative to $\beta_1$ is almost identical but the normalization factor of the non-unitary operation causes the last term to appear in the first equation. However, since this term is directly proportional to the variational state itself, no extra expectations value must be measured to take it into account. The number of expectation values that one must measure to estimate all the components of the gradient scales with the lattice size, one must take this into account because at some point, estimating the gradient through finite differences can become more efficient.

\vfill

\pagebreak
\newpage

\bibliography{manuscript}

\end{document}